\documentclass[12pt,reqno]{amsart}
\usepackage{amsmath}
\usepackage{fix-cm}
\usepackage{amsxtra}
\usepackage{amscd}
\usepackage{amsthm}
\usepackage{amsfonts}
\usepackage{amssymb}
\usepackage{mathtools}
\usepackage{eucal}
\usepackage{cases}
\usepackage{paralist} 
\usepackage{array}
\usepackage{longtable}
\usepackage{color}
\usepackage{float}
\usepackage[section]{placeins}
\usepackage{cite}
\usepackage{tikz} 
\usetikzlibrary{positioning}
\usetikzlibrary{decorations.pathreplacing}

\usepackage{graphicx,color}

\definecolor{darkblue}{rgb}{0,0,.5}
\definecolor{darkred}{rgb}{.5,0,0}
\definecolor{darkgreen}{rgb}{0,0.5,0}

\newcommand{\oa}{\mathbf{a}}
\newcommand{\oad}{\bar{\mathbf{a}}}

\usepackage[hang,flushmargin]{footmisc} 
\usepackage[arrow, matrix, curve]{xy}
\usepackage{perpage}
\MakePerPage{footnote}

\usepackage{hyperref}
\hypersetup {
	pdfstartview=FitH,
	colorlinks,
	citecolor=darkgreen,
	linkcolor=darkred,
	urlcolor=darkblue
}

\allowdisplaybreaks

\def\smallskip{\vskip\smallskipamount}
\def\medskip{\vskip\medskipamount}
\def\bigskip{\vskip\bigskipamount}

\newcommand{\ket}[1]{{| #1 \rangle}}      

\newcommand{\bea}{\begin{eqnarray}}
	\newcommand{\ena}{\end{eqnarray}}
\def\bel{\begin{eqnarray}}
	\def\enl{\end{eqnarray}}

\def\ba{\begin{eqnarray}}
	\def\ea{\end{eqnarray}}

\newcommand{\ID}{I}

\def\be{\begin{equation}}
	\def\ee{\end{equation}}

\newcommand{\C}{{\mathbb C}}
\newcommand{\Z}{{\mathbb Z}}

\newcommand{\id}{{\rm I}}

\newenvironment{tenumerate}{
	\begin{enumerate}
		
	}{\end{enumerate}}
\newcommand{\bi}{\begin{tenumerate}}
	\newcommand{\ei}{\end{tenumerate}}
\newcommand{\isoto}[1][]%
{{\mathop{\buildrel{\sim}\over\longrightarrow}\limits_{#1}}}

 \let\x=\xi \let\p=\pi \let\r=\rho

\def\2{\frac{1}{2}} \def\4{\frac{1}{4}}

\def\6{\partial}

\def\+{\dagger}

\def\<{\langle} \def\>{\rangle}

\def\i{{\rm i}}

\def\detq{{\rm det}_q}

\numberwithin{equation}{section}
\newtheorem{thm}{Theorem}[section]
\newtheorem{prop}[thm]{Proposition}

\newtheorem{rem}[thm]{Remark}
\newtheorem{cor}[thm]{Corollary}
\newtheorem{definition}[thm]{Definition}

\newcounter{saveenumi}

\begin{document}
\raggedbottom
	\begin{title}[]
		{Non-factorisable Yangian Lax matrices through oscillators}
	\end{title}
	\author{R.~Frassek, H.~Jürgens}
	\address{Math Department, University of Modena and Reggio Emilia, Via Campi 213b, 41125
		Modena, Italy}\email{henrik\_jue@live.de}\email{rfrassek@unimore.it}
	
\begin{abstract}
We construct oscillator realisations of Yangian Lax operators in
type $A$, with particular interest in non-factorisable operators
associated with prime snake modules. Our construction applies to
finite-dimensional skew representations of $Y(\mathfrak{gl}_n)$.
We prove that these representations are generated by the tensor
vacuum in suitably ordered tensor products of oscillator
representations. Their Lax operators are obtained by restricting
the corresponding products of linear oscillator Lax operators
to these submodules. We determine the vacuum eigenvalues and
Drinfeld polynomials, and give a determinant formula for the
dimensions directly in terms of the highest loop weights.
A $21$-dimensional example illustrates the construction,
including an explicit basis and a transfer-matrix identity.
\end{abstract}

	\maketitle
	
	\tableofcontents
		\section{Introduction}
		
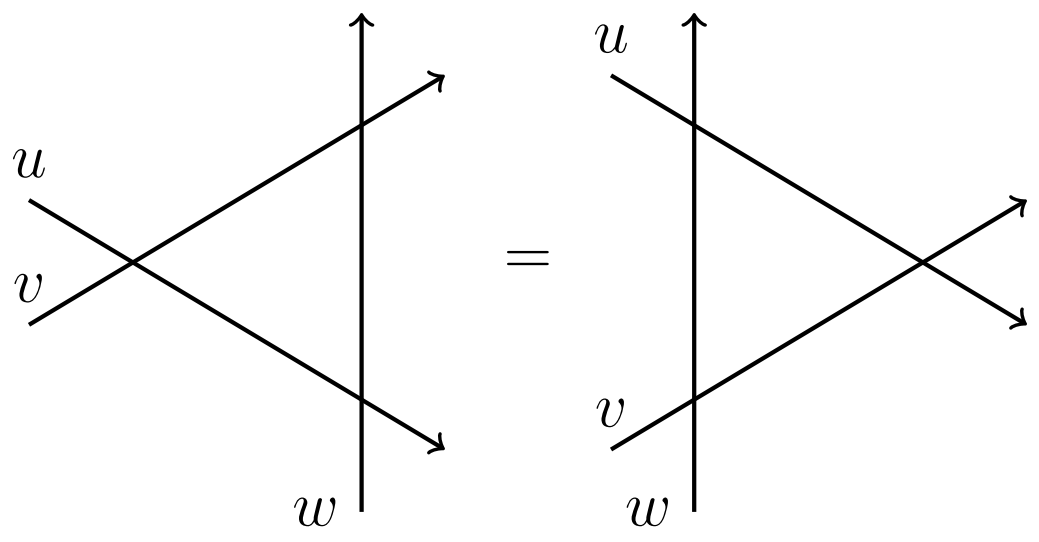
\begin{figure}[htbp]
			\centering
			
			\begin{tikzpicture}[scale=0.55]
				\draw [thick, ->] (3,0.25) node[anchor=east]{$w$} -- (3,6.25) ;
				\draw [thick, ->] (-1,2.5) node[anchor=south]{$v$} -- (4,5.5);
				\draw [thick, ->] (-1,4) node[anchor=south]{$u$} -- (4,1);
				\draw (5, 3.25) node{$=$};
				\draw [thick, ->] (7,0.25) node[anchor=east]{$w$} -- (7,6.25) ;
				\draw [thick, ->] (6,1) node[anchor=south]{$v$} -- (11,4);
				\draw [thick, ->] (6,5.5) node[anchor=south]{$u$} -- (11,2.5);
			\end{tikzpicture}
			
			\caption{The Yang--Baxter equation.}
			\label{fig:Yang--Baxter}
		\end{figure}

		Lax operators are the local building blocks of quantum integrable
		spin chains. In this paper, we construct explicit oscillator
		realisations of finite-dimensional irreducible Lax operators in
		type $A$, with particular interest in those which cannot be
		factorised into Lax operators on smaller quantum spaces.
		We consider $n\times n$ matrices $L(u)$ with entries acting on a
		quantum space $W$, satisfying the $RLL$ relation
		\begin{equation}
			R(u-v)(L(u)\otimes\id)(\id\otimes L(v))
			=(\id\otimes L(v))(L(u)\otimes\id)R(u-v),
			\label{eqn:RLL}
		\end{equation}
		where
		\begin{equation}
			R(u-v)=(u-v)\id+\operatorname{P}
			\in\operatorname{End}(\C^n\otimes\C^n)
			\label{eq:Rmat}
		\end{equation}
		is the rational $R$-matrix. Here $u,v\in\C$ are spectral
		parameters, $\id$ denotes the identity on the relevant space,
		and $\operatorname{P}(a\otimes b)=b\otimes a$ the permutation.
		The tensor products in \eqref{eqn:RLL} refer to the two auxiliary
		spaces; both copies of $L$ act on the same quantum space $W$.
		The $R$-matrix satisfies the Yang--Baxter equation
		\begin{equation}
			R_{1,2}(u-v)R_{1,3}(u-w)R_{2,3}(v-w)
			=R_{2,3}(v-w)R_{1,3}(u-w)R_{1,2}(u-v),
			\label{eq:YBE}
		\end{equation}
		where the subscripts specify the factors of
		$\C^n\otimes\C^n\otimes\C^n$ on which the matrices act;
		see Figure~\ref{fig:Yang--Baxter}.
		
		For a formal series $T(u)=\id+O(u^{-1})$, the relation
		\eqref{eqn:RLL} gives the RTT presentation of the Yangian
		$Y(\mathfrak{gl}_n)$ \cite{molev2007yangians}.
		An irreducible $Y(\mathfrak{sl}_n)$-representation determines its
		Lax operator up to scalar normalisation. This can be expressed
		through the (pseudo-)universal $R$-matrix
		$\mathcal R(u)\in
		(Y(\mathfrak{sl}_n)\otimes Y(\mathfrak{sl}_n))[[u^{-1}]]$,
		which intertwines the shifted coproduct and its opposite
		\cite{CPBook}. For an irreducible representation $\rho$ on
		$W_\rho$ and the fundamental evaluation representation $\pi$
		at parameter zero, the corresponding Lax operator has the form
		\begin{equation}
			L(u)=f(u)(\pi\otimes\rho)(\mathcal R(u))
			\in\operatorname{End}(\C^n)\otimes\operatorname{End}(W_\rho),
			\label{eqn:L_operator_irred}
		\end{equation}
		with a scalar factor $f(u)$. Multiplication of the RTT series by
		$h(u)\in1+u^{-1}\C[[u^{-1}]]$ generally changes the
		$Y(\mathfrak{gl}_n)$-action, while leaving its restriction to
		$Y(\mathfrak{sl}_n)$ unchanged; see
		Subsection~\ref{subsec:loop_weight_conventions}.
		
		Tensor products of Yangian modules correspond to ordered products
		of their Lax matrices, with the entries of each factor acting on
		its own quantum space. An irreducible $Y(\mathfrak{sl}_n)$-module
		is called prime if it cannot be written as a tensor product of
		two modules of dimension greater than one. Such modules therefore
		give non-factorisable irreducible Lax operators.
		For $Y(\mathfrak{gl}_2)$, every finite-dimensional irreducible
		Lax operator can be written as a product of linear evaluation
		Lax operators after scalar normalisation
		\cite[Section~4]{NT1998}. For $\mathfrak{gl}_3$ and higher,
		prime modules can instead require Lax operators of higher degree
		in the spectral parameter.
		
		We focus on the representations associated with skew Young
		diagrams \cite{NT1998,Nazarov2002}. Their
		$Y(\mathfrak{sl}_n)$-restrictions are the type $A$ snake modules
		of Mukhin and Young \cite{MY,MY2}. These modules have explicit
		$q$-characters described by families of non-overlapping paths.
		Nazarov and Tarasov constructed their Gelfand--Tsetlin bases in
		\cite{NT1998}, while Nazarov's fusion construction realises them
		as submodules of ordered tensor products of fundamental modules
		\cite{Nazarov2002}.
		
		Explicit Yangian realisations using Howe duality and Mickelsson
		algebras are developed in \cite{KhoroshkinNazarov2012}.
		Khoroshkin, Nazarov and Shapiro
		\cite[Sections~1.2 and~2.2--2.3]{KhoroshkinNazarovShapiro2014}
		realise irreducible modules as intertwiner images in tensor products
		of fundamental evaluation modules and identify their tensor-product
		highest-weight vectors. Our contribution is an explicit Lax
		construction for skew modules in terms of oscillator evaluation
		modules, see e.g. \cite{FKT,Derkachov:2008aq}. We decompose the skew diagram into disjoint
		shifted Young diagrams and associate one linear oscillator Lax
		operator with each block. The diagram fixes the weights and spectral
		shifts of these factors, while Nazarov's fusion fixes their order.
		Theorem~\ref{thm:Osc_Lax} shows that the tensor vacuum generates
		the corresponding finite-dimensional irreducible skew module under
		this ordered product. Restricting the product to that submodule
		and removing the common scalar factor gives its Lax operator with
		the prescribed lowest-weight eigenvalues. This restriction is
		non-factorisable when the skew module is prime.
		
		We also determine the Drinfeld polynomials and the diagonal
		vacuum eigenvalues directly from the skew diagram.
		The path description gives a dimension determinant directly in
		terms of the dominant loop weight for arbitrary $\mathfrak{gl}_n$. We work out the $21$-dimensional prime snake
		module of $Y(\mathfrak{gl}_3)$ in detail. It is irreducible under
		the Yangian but reducible under $\mathfrak{gl}_3$, and its Lax
		operator is quadratic and is not obtained from the inversion
		construction sketched below. We give explicit oscillator and Gelfand--Tsetlin
		bases, relating the loop weights to the eigenvalues of principal
		quantum minors and constructing the basis vectors with the
		corresponding ladder operators.

An explicit family of such modules occurs in the analysis of the
snail operator for the $\mathfrak{sl}_3$-invariant model
\cite[Section~4.1]{JuergensBoos2025}. We give its oscillator product
in Subsection~\ref{subsubsec:alternating_snakes}. We also briefly discuss
transfer matrices with the $21$-dimensional module in the quantum
or auxiliary space. The latter application connects the construction
to the known tableau and quantum Jacobi--Trudi formulas
\cite{Tsuboi1997,LuMukhin2021}.

	\subsection{Structure of the article}

In Section~\ref{sect:gen_def} we fix the Yangian and spectral
conventions and recall loop weights, $q$-characters, snake modules
and their description by paths and skew diagrams. We then relate
loop weights and basis vectors to the Gelfand--Tsetlin algebra.
Section~\ref{sect:snake/skew_lax} develops the oscillator construction.
We first determine the vacuum eigenvalues and dimensions, then
decompose a general skew diagram into Young blocks and prove the
main theorem. The examples include the $21$-dimensional prime
snake module and an alternating family of arbitrary length.
We conclude the section with applications to transfer matrices.
Section~\ref{sect:conclusion} summarises the results and open questions.
Appendix~\ref{app:21dim_module} contains the explicit oscillator and
Gelfand--Tsetlin bases and Appendix~\ref{app:skew_dimension_determinant}
displays the dimension determinant for the $\mathfrak{gl}_3$ diagrams.

	\subsection*{Acknowledgement}

	We thank our friends and colleagues Frank Göhmann, Artur Hutsalyuk, Alexander Razumov and Sasha Tsymbaliuk for helpful discussions and advice during the research and writing of this paper.
	All authors were supported by the grant FAR UNIMORE project CUP-E93C23002040005. RF was supported by INdAM (GNFM), INFN and by the PRIN project CUP-E53D23002220006.

	\section{General definitions}
	\label{sect:gen_def}
    In type $A$, $n$ denotes the dimension of the auxiliary vector space,
    while $\ell=n-1$ is the rank and $I=\{1,\ldots,\ell\}$ is the set of
    Dynkin nodes.
		\subsection{The linear Lax (Evaluation homomorphism)}
        \label{subsec:eval_conventions}
		A well-known solution to the RLL equation \eqref{eqn:RLL} is linear in the spectral parameter and of the form
	\begin{equation}
		L(u)=u\id+\sum_{a,b=1}^n e_{ab}J_{ba}\,.
        \label{eqn:lin_Lax}
	\end{equation}
This solution arises from the evaluation homomorphism from $Y(\mathfrak{gl}_n)$ to the universal enveloping algebra $U(\mathfrak{gl}_n)$. Here, we introduced the unit matrices $e_{ab}$ with $(e_{ab})_{cd}=\delta_{ac}\delta_{bd}$ and the generators of $\mathfrak{gl}_n$   with commutation relations
	\begin{equation}
		[J_{ab},J_{cd}]=\delta_{bc}J_{ad}-\delta_{da}J_{cb}.
	\end{equation}
	Finite-dimensional irreducible representations of $\mathfrak{gl}_n$ can be labelled by the eigenvalues of the Cartan elements $J_{aa}$ on the highest-weight state $|hws\rangle$
	\begin{equation}
	J_{ab}|hws\rangle =\begin{cases}
	  0 &a<b\\
	  \lambda_a |hws\rangle  & a=b
	 \end{cases}\,,
	\end{equation}
	where $\lambda=(\lambda_1,\ldots,\lambda_n)\in\C^n$ is dominant integral,
    meaning $\lambda_i-\lambda_{i+1}\in\Z_{\geq0}$ for $i\in I$.
    The common additive part of the $\lambda_i$ is unrestricted. The unit
    matrices $e_{ab}$ arise from the first fundamental representation of
    $\mathfrak{gl}_n$.

    Because the generators enter $L_{ab}(u)$ as $J_{ba}$, this Lie
    highest-weight vector is annihilated by $L_{ab}(u)$ for $a>b$.
    We call it a Yangian lowest-weight vector in this matrix convention.
    Its dominant loop weight is nevertheless called the highest loop
    weight in the $q$-character terminology used below.

    \medskip

 \paragraph{\textit{Evaluation parameters and scalar normalisation.}}
    \leavevmode\par
   For the evaluation representation on $V_\lambda$ with inhomogeneity $z$,
we shift the argument of the linear Lax operator by $z$ and divide by
$u-z$
    \begin{equation}
        T_{\lambda,z}(u)
        =\frac{L_\lambda(u-z)}{u-z}
        =\id+\frac{\sum_{a,b}e_{ab}J_{ba}^{(\lambda)}}{u-z}.
        \label{eqn:evaluation_RTT_convention}
    \end{equation}
    The coefficients of the expansion of $T_{\lambda,z}(u)$ in
    $u^{-1}$ give the action of the Yangian generators. A spectral
    translation acts as $\tau_z:T(u)\mapsto T(u-z)$. 
    This agrees with the positive-sign evaluation convention of
    \cite{NT1998,FKT}.

    If $\lambda=\nu+c(1,\ldots,1)$, identify the underlying vector spaces
    so that $J_{ab}^{(\lambda)}=J_{ab}^{(\nu)}+c\delta_{ab}\id$.
    Then
    \begin{align}
        L_\lambda(u-z)&=L_\nu\bigl(u-(z-c)\bigr),
        \label{eqn:weight_shift_polynomial_Lax}\\
        T_{\lambda,z}(u)&=
        \frac{u-z+c}{u-z}\,T_{\nu,z-c}(u).
        \label{eqn:weight_shift_RTT_twist}
    \end{align}
   Thus a common shift of the weights $\lambda_i$ can be absorbed
into the inhomogeneity $z$. When comparing the
    $Y(\mathfrak{gl}_n)$-representations defined by $T(u)$, we must
    also keep the scalar factor in \eqref{eqn:weight_shift_RTT_twist}.

    Nazarov's convention in \cite{Nazarov2002} has evaluation matrix
    $\id-\sum_{a,b}e_{ab}J_{ba}/(x-z_{\mathrm{Naz}})$.
    To use it here we put $x=-u$, so that
    $z=-z_{\mathrm{Naz}}$. This is a change of spectral variable;
    the representation is the same.

Besides the linear solutions \eqref{eqn:lin_Lax}, there exist solutions of higher order in the spectral parameter. This follows by multiplying the RLL equation by inverse Lax operators. After reversing the spectral parameter, we find that $\bar L(u)=L^{-1}(-u)$ is again a solution of the RLL equation. Writing $M\coloneqq\sum_{a,b=1}^n e_{ab}J_{ba}$, its expansion at infinity is
	\begin{equation}
		L^{-1}(-u)= -\sum_{k=0}^{\infty}\frac{M^k}{u^{k+1}},
	\end{equation}
	which is an identity of formal series in $u^{-1}$. On a
    finite-dimensional representation,
    $(-u\id+M)^{-1}$ is a rational matrix function and hence meromorphic
    on the Riemann sphere. For example, if $M^2=M$, then
    \begin{equation*}
        L^{-1}(-u)=-\frac{\id-M}{u}-\frac{M}{u-1}.
    \end{equation*}
    On an irreducible $V_\lambda$, a finite expression in powers of $M$
    can be obtained using the characteristic identity
	\begin{equation}
	 \prod_{k=1}^n (M-\ell_k \id)=0\,,
	\end{equation}
	with shifted weights $\ell_k=\lambda_k-k+1$, see \cite{Green:1971rp}.

The coproduct structure of the Yangian can then be used to construct further solutions by multiplying Lax matrices in the $n$-dimensional space and taking tensor products in $V_\lambda$. These solutions are also of higher order in the spectral parameter, but need not be Yangian irreducible; see \cite{Molev00} for irreducibility criteria for tensor products of evaluation modules.

We are interested in prime irreducible Yangian representations whose
restriction to $\mathfrak{gl}_n$ is reducible. Their Lax operators admit
neither a proper non-zero invariant subspace in the quantum space nor a
factorisation corresponding to a tensor product of two Yangian modules
of dimension greater than one.
For this purpose, we introduce loop weights and Drinfeld polynomials.

    \subsection{\texorpdfstring{$l$}{l}-weights, q-characters and the Gauß LDU decomposition of the \texorpdfstring{$L$}{L}-operator}
	\label{subsec:loop_weight_conventions}
    Let $L(u)$ be the Lax operator of a finite-dimensional Yangian
    representation $\rho$. We apply a Gauss $LDU$ decomposition of the form
	\begin{equation}
		L(u) = L^-(u)\, D(u) \, L^+(u),
	\end{equation}
	where $L^-(u)$ and $L^+(u)$ are lower and upper unitriangular,
    respectively. We denote the diagonal entries of $D(u)$ by
    $D_i(u)$, without shifting their arguments. For example,
    \begin{equation*}
        D_1(u)=L_{11}(u),\qquad
        D_2(u)=L_{22}(u)-L_{21}(u)L_{11}(u)^{-1}L_{12}(u).
    \end{equation*}
    Their consecutive ratios define the loop-weight eigenvalues. On a
    common eigenvector $\xi_k$ we write
	\begin{equation}
		D_{i+1}(u)/D_i(u) \cdot \xi_k \eqqcolon \psi_{i,k}(u)\cdot \xi_k ,\quad i\in I.\label{eqn:Gauß_dec1}
	\end{equation}
	These eigenvalues are parametrised by rational functions $\nu_{i,k}(u) \coloneqq \frac{Q_{i,k}(u)}{R_{i,k}(u)}$ via
	\begin{equation}
		\psi_{i,k}(u) = \frac{\nu_{i,k}(u-1)}{\nu_{i,k}(u)} = \frac{Q_{i,k}(u-1)}{Q_{i,k}(u)}\frac{R_{i,k}(u)}{R_{i,k}(u-1)}\label{eqn:Gauß_dec2}.
	\end{equation}
	The $\nu_{i,k}$ are called the loop-weights {($l$-weights)} of the representation $\rho$ and the $Q_{i,k}$ (resp. $R_{i,k}$) the dominant (resp. antidominant) part of $\nu_{i,k}$ (see \cite{NT1998}, \cite{FR}, \cite{FM}, or \cite{JuergensMaster}). We take $Q_{i,k}$ and $R_{i,k}$ to be coprime and monic, which fixes this rational representation uniquely.\\
	For $k=1$, let $\xi_1 = \xi$ be the
    lowest-weight vector of $\rho$, i.e. $\xi$ is annihilated by all $L_{ab}(u)$ with $a>b$. Then, the fundamental observation is that $R_{i,1} = 1$ and $Q_{i,1} = P_i$. In this case, the $l$-weight $\nu_{i,1}$ is called dominant, and $P_i$ is the $i$-th Drinfeld polynomial of the representation $\rho$.
    On this lowest-weight vector, \eqref{eqn:Gauß_dec1} simplifies to
    \cite{Drinfeld1988,NT1998}:
	\begin{equation}
		\frac{L_{i+1,i+1}(u-i)}{L_{i,i}(u-i)}\,\xi = \frac{P_i(u-1)}{P_i(u)}\,\xi,\quad i\in I \label{eqn:Shifted_L_ii_to_Drin_polys_direct}.
	\end{equation}
    For $Y(\mathfrak{sl}_n)$, every collection of monic polynomials
    $P_i(u)$, $i=1,\ldots,n-1$, determines a unique finite-dimensional
    irreducible representation up to isomorphism, and every finite-dimensional
    irreducible representation arises in this way \cite{FR,FM,CPBook,NT1998}.
    For $Y(\mathfrak{gl}_n)$, the Drinfeld polynomials determine the
    irreducible representation only up to the scalar automorphisms
    \begin{equation}
        \omega_f:T(u)\longmapsto f(u)T(u),
        \qquad f(u)\in1+u^{-1}\C[[u^{-1}]].
        \label{eqn:scalar_Yangian_twist}
    \end{equation}
    These act trivially on $Y(\mathfrak{sl}_n)$ \cite{NT1998,Nazarov2002}.
    Thus a dominant monomial determines the irreducible
    $Y(\mathfrak{sl}_n)$-representation. To specify the full
    $Y(\mathfrak{gl}_n)$-action, we also fix the scalar factor in
    $T(u)$. Multiplying $T(u)$ by $f(u)$ does not change its
    invariant subspaces: the coefficients of $f(u)T(u)$ are linear
    combinations of those of $T(u)$, and conversely by using $f(u)^{-1}$.

    The common scalar factor cancels from the ratios in
    \eqref{eqn:Gauß_dec1}. Below, we fix the polynomial eigenvalues
    of $L_{aa}(u)$ on the lowest-weight vector by requiring them
    to be monic and to have greatest common divisor $1$.

    It is convenient to introduce the fundamental loop weights
    $Y_{i,a}$ by the following identification in the $i$-th component:
    \begin{equation}
        Y_{i,a}= u+\frac{a-i+1}{2},
        \qquad i\in I,\quad a\in\C.
        \label{eqn:fund_l-weights}
    \end{equation}
    More precisely, let $\varphi_j(Y_{i,a})(u)=(u+(a-i+1)/2)^{\delta_{ji}}$. If
    $m=\prod_{i,a}Y_{i,a}^{u_{i,a}}$, with finitely many non-zero integer
    exponents, its rational loop-weight components are
    \begin{equation*}
        \nu_j(u)=\varphi_j\!\left(\prod_{i,a}Y_{i,a}^{u_{i,a}}\right)(u)
        =\prod_a\left(u+\frac{a-j+1}{2}\right)^{u_{j,a}}.
    \end{equation*}
    Thus $Y_{i,a}$ corresponds to a zero at $u=-(a-i+1)/2$.
    This choice agrees with \eqref{eqn:Shifted_L_ii_to_Drin_polys_direct}
    and the $q$-characters below. By abuse of notation, we also call
    $m$ the loop weight and, when it is dominant, identify it with the corresponding
    collection of Drinfeld polynomials. Hence, the collection of {eigenvalues} of the Cartan part of the Yangian is encoded in loop weights, and we can define the $q$-character of a representation $V$ as the {Laurent polynomial} $\chi_q(V)\in \mathbb{Z}[Y_{i,a}^{\pm 1}]_{i\in I, a\in\mathbb{C}}$ in the fundamental loop weights given by
	\begin{equation}
		\chi_q(V)\coloneq\sum_{W}\dim(W) m(W),
	\end{equation}
	where the sum runs over the joint generalised eigenspaces $W$
	of the coefficients of $D_{i+1}(u)/D_i(u)$, called loop-weight
	spaces, and $m(W)$ denotes the corresponding monomial.

    We call a {loop}{-weight} $m = \prod_{i\in I, a\in\mathbb{C}}Y_{i,a}^{u_{i,a}}$ \textit{dominant} if it is a {monomial} in the {fundamental loop weights}, i.e. $u_{i,a}\geq 0$. In the following, assume that $\mathfrak{g}$ is {simply laced} (in particular, of type $A$), and let $\operatorname{wt}:\mathbb{Z}[Y_{i,a}^{\pm 1}]_{i\in I, a\in\mathbb{C}}\to \mathbb{Z}[P]$ be the homomorphism which maps fundamental Yangian weights onto fundamental Lie-algebra weights, i.e. $Y_{i,a}\mapsto e_{\omega_i}$.\footnote{That is, we forget the affine parameter $a\in\mathbb{C}$.} In the additive loop parameters used here, we define {affine simple roots}
		\begin{equation*}
			A_{i,a} = Y_{i,a-1}Y_{i,a+1}\prod_{j\neq i}Y^{a_{ij}}_{j,a},
		\end{equation*}
	such that $\operatorname{wt}(A_{i,a}) = e_{\alpha_i}$, where
    $A=(a_{ij})$ is the Cartan matrix of $\mathfrak{g}$.
    In type $A_\ell$, this is
    $A_{i,a}=Y_{i,a-1}Y_{i,a+1}Y_{i-1,a}^{-1}Y_{i+1,a}^{-1}$,
    with boundary factors omitted. The spectral translation $\tau_z$
    of Subsection~\ref{subsec:eval_conventions} sends
    \begin{equation}
        P_i(u)\longmapsto P_i(u-z),
        \qquad Y_{i,a}\longmapsto Y_{i,a-2z}.
        \label{eqn:loop_spectral_translation}
    \end{equation}
    A change of one unit in the spectral variable therefore corresponds
    to two units in the loop label.
	
	Then, the {affine root lattice} $\boldsymbol{\mathcal{Q}}$ is defined as the group of  {Laurent monomials in} the {affine simple roots} and we denote by $\boldsymbol{\mathcal{Q}^\pm}$ the  {monomials} in the $A_{i,a}^{\pm1}$, respectively.
	Consequently, we define a {partial order} on loop weights by $m\leq m'$ if $m'm^{-1}\in\mathcal{Q}^+$. This order is compatible with the partial order on the {weight lattice} $P$: if $\operatorname{wt}(m)=e_\lambda$ and $\operatorname{wt}(m')=e_{\lambda'}$, then $m\leq m'$ implies $\lambda\leq\lambda'$.
	In this case, we say "$m'$ {is} {higher than} $m$".

    With this terminology, the highest-loop-weight theorem can be formulated as follows; see \cite[Section~3.2]{GT} and \cite{JuergensMaster,HL,FM,FR}.
	\begin{thm}[Theorem of the highest loop weight]\label{thm:highest_loop_weight}
		Let $Y(\mathfrak{g})$ be the Yangian of the simple Lie algebra $\mathfrak{g}$. Then
		\begin{enumerate}
			\item every {irreducible finite-dimensional representation} has a {highest loop-weight},
			\item the {highest loop-weight} is always  {dominant} and usually called the  {Drinfeld polynomial} of the representation,
			\item two {irreducible representations} with the {same highest loop-weight} are {isomorphic},
			\item every {dominant} {loop-weight} $\boldsymbol{m}$ (or {Drinfeld polynomial}) is the {highest loop-weight} of an {irreducible representation} $V(\boldsymbol{m})$. $\odot$
		\end{enumerate}
	\end{thm}
     
	\subsection{Snake-modules and the path formula}
	For the integral spectral subcategory introduced below, we use
the correspondence with quantum loop algebra representations of the
same type for transcendental $q$. This correspondence is compatible
with $q$-characters \cite[Sections~1.8 and~7.7]{GT}, allowing us to
use the quantum affine $q$-character formulas in the Yangian setting.
Let us briefly recall the so-called snake modules
\cite{MY,MY2,CPBook}. As we shall see, this class of representations admits
a particularly explicit description in terms of skew diagrams and has a
number of useful structural properties.

For later use, we recall the following terminology.
\begin{definition}[special, thin, prime, real]~
	\begin{enumerate}
		\item A representation $V\in\operatorname{Rep}(Y)$ is called {special} if $\chi_q(V)$ has {exactly one dominant loop-weight}.
		\item $V$ is called {thin} if {no loop-weight space} of $V$ has  {dimension greater} than  {one}.
		\item $V$ is called {prime} if it is {not isomorphic to a tensor product} of two representations of dimension greater than one.
		\item $V$ is called {real} if $V\otimes V$ is {irreducible}. $\odot$
	\end{enumerate}
\end{definition}
In types $A$ and $B$, the $q$-characters of a large class of irreducible representations can be computed by the path formula for snake modules \cite{MY}; in type $A$, see also \cite{NT1998}. We briefly recall this construction.

Define the {subset} $$\mathcal{X}:=\{(i,k)\in I\times\mathbb{Z}:i-k=1\mod 2\}\,.$$
We first work with representations whose $q$-characters lie in
$\Z[Y_{i,k}^{\pm1}]_{(i,k)\in\mathcal X}$. These form a subcategory
$\mathcal C_{\Z}$ of
$\mathcal C=\operatorname{Rep}(Y(\mathfrak{sl}_n))$
closed under taking tensor products.

Conversely, every simple object $S$ of $\mathcal C$ can be written
as a tensor product
\begin{equation*}
	S \cong S_1(a_1)\otimes\cdots\otimes S_k(a_k)\,,
\end{equation*}
for some simple objects $S_1,\ldots,S_k\in\mathcal C_{\Z}$ and
parameters $a_1,\ldots,a_k\in\C$ satisfying
$a_r-a_s\notin2\Z$ for $r\neq s$ \cite{CPBook,C2002}.
Here $S_r(a_r)$ denotes the pullback of $S_r$ by the spectral
translation automorphism $\tau_{-a_r/2}$. By
\eqref{eqn:loop_spectral_translation}, its $q$-character is obtained
from that of $S_r$ by replacing every $Y_{i,k}$ with $Y_{i,k+a_r}$.
The condition on the parameters means that the Drinfeld roots
of different factors belong to distinct classes modulo $\Z$.

{Snake modules} are then defined as follows.
Let $(i,k)\in\mathcal{X}$. A point $(i',k')$ is said to be in {\textit{snake position}} with respect to $(i,k)$ iff
\begin{equation}
	k'-k \geq |i'-i|+2.\label{eqn:snake_pos}
\end{equation}
The point $(i',k')$ is in \textit{minimal snake position} with respect to $(i,k)$ iff $k'-k$ is equal to the {lower bound}.
Following \cite[Section~3.2]{DuanLiLuo2019}, we say that $(i',k')\in\mathcal{X}$ is in {prime snake position} with respect to $(i,k)$ iff
\begin{equation}
	\min(2\ell+2-i-i',i'+i)\geq k'-k \geq |i'-i|+2.\label{eqn:prime_snake_pos}
\end{equation}
Let $M\in\mathbb{N}$. A finite sequence $(i_t,k_t)_{1\leq t\leq M}$ of points in $\mathcal{X}$ is a {\textit{snake}} iff for all $2\leq t\leq M$, $(i_t,k_t)$ is in \textit{snake position} with respect to $(i_{t-1},k_{t-1})$.
It is a {\textit{minimal}} (resp. {\textit{prime}}) {\textit{snake}} iff any two successive points are in \textit{minimal} (resp. \textit{prime}) \textit{snake position} with respect to each other.
The simple module $\boldsymbol{V(m)}$ is a ({\textit{minimal/prime}}) \textit{{snake module}} iff $\boldsymbol{m} = \prod_{t=1}^{M}Y_{i_t,k_t}$ for some (\textit{minimal/prime}) \textit{snake} $(i_t,k_t)_{1\leq t\leq M}$. These variants are distinguished by the properties they imply.

{\textit{Snake modules}} are  {special} and {thin}.
A {\textit{snake module}} is {prime} iff its \textit{snake} is \textit{{prime}}
\cite[Theorem~3.3]{DuanLiLuo2019}, and
{\textit{prime snake modules}} are {real}
\cite[Theorem~3.4]{DuanLiLuo2019}.

Equivalently, whenever a {Drinfeld polynomial} $\boldsymbol{m}$ has zeros forming a snake, the corresponding irreducible module $\boldsymbol{V(m)}$ is called a {snake module}. Moreover, the $q$-character of these modules admits a simple description in terms of paths on a square lattice: For every linear factor in the Drinfeld polynomial, we define a corresponding starting point and endpoint on the lattice. Every monomial in the $q$-character then corresponds to exactly one family of non-overlapping {diagonal paths} connecting all starting points and endpoints. Thus, the entire $q$-character is obtained by summing over all such non-overlapping path families.

To make this precise, we can define a {set} $\mathcal{P}_{i,k}$ {of} {paths} of type $A_\ell$ by
\begin{equation*}
	\begin{split}
		\mathcal{P}_{i,k}\coloneq \{&((0,y_0),(1,y_1),\dots,(\ell+1,y_{\ell+1}))|\\
		&y_0=i+k,\, y_{\ell+1}=\ell+1-i+k,\, y_{j+1}-y_j=\pm1\ (j=0,\ldots,\ell)\},
	\end{split}
\end{equation*}
and a mapping $$p\mapsto m(p)\coloneq \prod_{(j,r)\in C_{p,-}}Y_{j,r}\prod_{(j,s)\in C_{p,+}}Y_{j,s}^{-1}$$ from paths to loop weights.\footnote{We can also define {raising} and {lowering moves} for a path through the {action} of the {affine simple roots}.}
The sets $C_{p,+}$ and $C_{p,-}$ of upper and lower corners of $p\in\mathcal{P}_{i,k}$, respectively, are defined with respect to the $y$-coordinate by
\begin{equation*}
    \begin{aligned}
        C_{p,+}&\coloneqq\{(j,y_j):1\leq j\leq\ell,\ y_{j-1}=y_j-1=y_{j+1}\},\\
        C_{p,-}&\coloneqq\{(j,y_j):1\leq j\leq\ell,\ y_{j-1}=y_j+1=y_{j+1}\}.
    \end{aligned}
\end{equation*}
Then, for any {snake} $(i_t,k_t)_{1\leq t\leq T}$, we define $\overline{\mathcal{P}}_{(i_t,k_t)_{1\leq t\leq T}}$ as a set of $T$-tuples $(p_1,\dots,p_T)$ of {paths} $p_t\in \mathcal{P}_{i_t,k_t}$ which are \textit{non-overlapping}: writing
$p_a=((x,y_a(x)))_{x=0}^{n}$, we require
$y_a(x)<y_b(x)$ whenever $a<b$, for every $x=0,\ldots,n$.
Finally, the path formula reads as follows (see \cite{MY2}).
\begin{thm}[The path formula]
	Let $(i_t,k_t)_{1\leq t\leq T}$ be a snake in $\mathcal{X}$. Then
	\begin{equation*}
		\chi_q\left(V\left(\prod_{t=1}^{T}Y_{i_t,k_t}\right)\right) = \sum_{(p_1,\dots,p_T)\in\overline{\mathcal{P}}_{(i_t,k_t)_{1\leq t\leq T}}}\prod_{t=1}^T m(p_t).
	\end{equation*}
	The module $V\left(\prod_{t=1}^{T}Y_{i_t,k_t}\right)$ is {thin} and {special} (and anti-special). $\odot$
	\label{thm:path_formula}
\end{thm}

We next recall how the same snake data are encoded by Drinfeld polynomials and skew diagrams.

    \subsection{Drinfeld polynomials, snakes and skew diagrams}
	Let $S=(i_t,k_t)_{1\leq t\leq T}\subset\mathcal X$ be a snake in type $A_\ell$ \cite{MY,MY2} \cite{JuergensMaster}. We obtain the Drinfeld polynomials of the corresponding Yangian representation using \eqref{eqn:fund_l-weights} as
\begin{equation}
	P_i(u) = \prod_{t:i_t=i}\left(u+\frac{k_t-i+1}{2}\right),\quad i\in I.\label{eqn:DrinPoly_snake}
\end{equation}
Moreover, we can associate to it a skew diagram via \cite{NT1998,MY2}
\begin{equation}
	(\lambda,\mu)_{S}\coloneqq \left\{(s,t)\in \mathbb{Z}\times\mathbb{Z}_{\geq 0}:\,1\leq t\leq T,\,t-\frac{k_t+(i_t-1)}{2}\leq s \leq t-\frac{k_t-(i_t-1)}{2}\right\}.\label{eqn:snake_skew}
\end{equation}
Here $s$ is the row coordinate, increasing downwards, and $t$ is the
column coordinate, increasing to the right. The bottom box in column $t$
is at $s=t-(k_t-i_t+1)/2$, so its content is
\begin{equation*}
    c_t^{\mathrm{bottom}}=t-s=\frac{k_t-i_t+1}{2}.
\end{equation*}
This explains the sign in \eqref{eqn:DrinPoly_snake}: in the convention
of \cite[Eq.~(2.6)]{NT1998}, $P_i(u)$ is the product of $u+c$ over
bottom-box contents of columns of height $i$.
Thus, as in the Young diagram case, the column height gives the Dynkin node.
Translating the whole diagram by $s\mapsto s+r$, $t\mapsto t+h$
changes every content by $h-r$ and hence sends $P_i(u)$ to $P_i(u+h-r)$.
Translating the diagram therefore shifts the spectral parameter.
Since the other fundamental modules can be obtained from the first fundamental one, the corresponding module can be constructed by fusing fundamental $R$-matrices.
Using \eqref{eqn:Shifted_L_ii_to_Drin_polys_direct} we obtain for the lowest weight vector $\xi$
\begin{equation}
	\frac{L_{i+1,i+1}(u-i)}{L_{i,i}(u-i)}\,\xi = \prod_{t:i_t=i}\frac{u-1+\frac{k_t-i+1}{2}}{u+\frac{k_t-i+1}{2}}\,\xi,\quad i\in I.\label{eqn:Shifted_L_ii_to_Drin_polys}
\end{equation}
Before turning to the Gelfand--Tsetlin description of the same representations, it is useful to illustrate the path formula and the snake/skew correspondence in a few simple examples.

\subsection{Examples}
We illustrate the path formula and its relation to skew representations in the simplest non-trivial examples. We work in type $A_2$, i.e. with $\mathfrak{gl}_3$.

We are particularly interested in representations which are irreducible as Yangian modules but reducible as $\mathfrak{gl}_3$-modules, since these give rise to irreducible Lax operators which are not obtained directly through the evaluation homomorphism. We will arrive at a first prime example of this type by successive fusion of fundamental representations.

To find short exact sequences, we use the fact that fundamental
modules are snake modules and that snake modules are special
in type $A$. Thus, we can calculate their $q$-characters using
the path formula, multiply them, and check whether all dominant
monomials correspond to snake modules. If this is the case,
the theorem of the highest loop weight, together with the
specialness of snake modules, determines the irreducible
composition factors and their multiplicities. In the examples
below, we obtain two irreducible composition factors and hence
a short exact sequence. We first record the corresponding
identity in the Grothendieck ring.

\begin{figure}[htbp]
	\centering
	
	\begin{tikzpicture}[
		x=0.66cm,
		y=0.66cm,
        scale=1.25,
		line cap=round,
		line join=round
		]
		
		
		\node[anchor=west,font=\large] at (-5.35,-0.55)
		{$V(Y_{1,0}Y_{2,3})$};
		
		\node[anchor=west,font=\large] at (-5.35,-3.15)
		{Paths:};

		
		\node[anchor=east,font=\large] at (-1.25,0.12)
		{$A_2$};
		
		\node at (1,0.58) {$1$};
		\node at (2,0.58) {$2$};
		
		\foreach \r in {0,...,6}{
			\node[anchor=east] at (-0.55,-\r) {$\r$};
		}
		
		\foreach \x in {0,1,2,3}{
			\foreach \r in {0,...,6}{
				\fill (\x,-\r) circle (0.50pt);
			}
		}
		
		\draw[line width=0.75pt]
		(1,0)--(0,-1)--(1,-2)--(2,-1)--cycle;
		
		\draw[line width=0.75pt]
		(2,-1)--(1,-2)--(2,-3)--(3,-2)--cycle;
		
		\draw[line width=0.75pt]
		(2,-3)--(1,-4)--(2,-5)--(3,-4)--cycle;
		
		\draw[line width=0.75pt]
		(1,-4)--(0,-5)--(1,-6)--(2,-5)--cycle;
		
		\foreach \p in {
			(1,0),(0,-1),(2,-1),(1,-2),(3,-2),
			(2,-3),(1,-4),(3,-4),(0,-5),(2,-5),(1,-6)
		}{
			\fill \p circle (1.15pt);
		}

		
		\node[anchor=west,font=\normalsize] at (5.2,0.50)
		{``all paths $-$ overlapping paths''};
		
		
		\node[font=\fontsize{27}{27}\selectfont] at (5.05,-1.30) {$($};
		
		\draw[line width=0.72pt]
		(5.60,-0.95)--(6.35,-1.70)--(7.10,-0.95);
		
		\node[font=\normalsize] at (7.58,-1.30) {$+$};
		
		\draw[line width=0.72pt]
		(8.00,-1.05)--(8.60,-1.65)--(9.20,-1.05)--(9.80,-1.65);
		
		\node[font=\normalsize] at (10.25,-1.30) {$+$};
		
		\draw[line width=0.72pt]
		(10.70,-0.90)--(11.45,-0.15)--(12.30,-0.90);
		
		\node[font=\fontsize{27}{27}\selectfont] at (12.85,-1.30) {$)$};
		\node[font=\normalsize] at (13.30,-1.30) {$\cdot$};

		
		\node[font=\fontsize{27}{27}\selectfont] at (5.05,-3.30) {$($};
		
		\draw[line width=0.72pt]
		(5.60,-2.95)--(6.35,-3.70)--(7.10,-2.95);
		
		\node[font=\normalsize] at (7.58,-3.30) {$+$};
		
		\draw[line width=0.72pt]
		(7.95,-3.05)--(8.60,-2.40)--(9.25,-3.05)--(9.90,-2.40);
		
		\node[font=\normalsize] at (10.25,-3.30) {$+$};
		
		\draw[line width=0.72pt]
		(10.70,-2.95)--(11.55,-2.10)--(12.40,-2.85);
		
		\node[font=\fontsize{27}{27}\selectfont] at (12.85,-3.30) {$)$};
		
		\node[font=\normalsize] at (13.45,-3.30) {$-$};
		\node[font=\Large] at (14.55,-3.30) {$\Rightarrow$};

		
		\node[font=\large] at (5.95,-5.05) {$-$};
		
		\draw[line width=0.72pt]
		(6.85,-4.35)--(7.60,-5.10)--(8.35,-4.35);
		
		\draw[line width=0.72pt] (7.60,-5.10) circle (0.18);
		
		\node[font=\large] at (8.85,-4.90) {$\cdot$};
		
		\draw[line width=0.72pt]
		(9.70,-5.35)--(10.45,-4.60)--(11.20,-5.35);
		
		\draw[line width=0.72pt] (10.45,-4.60) circle (0.18);

		
		\node[anchor=west,font=\large] at (-5.35,-7.10)
		{
			$\chi_q\!\left(V(Y_{1,0}Y_{2,3})\right)
			=
			\left(
			Y_{1,0}
			+
			Y_{1,2}^{-1}Y_{2,1}
			+
			Y_{2,3}^{-1}
			\right)
			\left(
			Y_{2,3}
			+
			Y_{1,4}Y_{2,5}^{-1}
			+
			Y_{1,6}^{-1}
			\right)
			-1$
		};

		
		\node[anchor=west,font=\large] at (0.95,-8.45)
		{$\Rightarrow\ 8\ \mathrm{dim.\ adj.\ repr.}$};

		
		\node[font=\large] at (7.75,-8.25) {``};
		
		\begin{scope}[shift={(8.20,-8.58)},scale=0.5]
			\draw (0,0) rectangle (1,1);
			\draw (1,0) rectangle (2,1);
			\draw (0,-1) rectangle (1,0);
		\end{scope}
		
		\node[font=\normalsize] at (9.45,-8.25) {''};
		
	\end{tikzpicture}
	
	\caption{
		Path representation of $V(Y_{1,0}Y_{2,3})$
		with q-character in \eqref{qchar8}.
	}
	\label{fig:adj_qChar}
	
\end{figure}
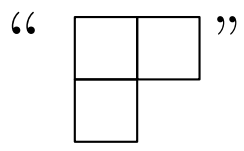

By Theorem~\ref{thm:path_formula}, the fundamental representation $V(Y_{1,0})$ has $q$-character
\begin{equation}\label{qch10}
	\chi_q(V(Y_{1,0})) = Y_{1,0}+Y_{1,2}^{-1}Y_{2,1}+Y_{2,3}^{-1}
\end{equation}
and the antifundamental representation $V(Y_{2,3})$ has $q$-character
\begin{equation}\label{qch23}
	\chi_q(V(Y_{2,3})) = Y_{2,3}+Y_{1,4}Y_{2,5}^{-1}+Y_{1,6}^{-1},
\end{equation}
as shown in Figures~\ref{fig:fund_qChar} and~\ref{fig:antifund_qChar}, respectively.
\begin{figure}[htbp]
	\centering
	
	\begin{tikzpicture}[
		x=0.85cm,
		y=0.85cm,
		baseline=(current bounding box.center)
		]
		
		
		\node[anchor=west] at (-5.2,-0.55) {\large $V(Y_{1,0})$};
		\node[anchor=west] at (-5.2,-2.05) {\large Paths:};

		
		\node[anchor=east] at (-1.15,0.15) {$A_2$};
		
		\node at (1,0.55) {$1$};
		\node at (2,0.55) {$2$};
		
		\node[anchor=east] at (-0.55, 0) {$0$};
		\node[anchor=east] at (-0.55,-1) {$1$};
		\node[anchor=east] at (-0.55,-2) {$2$};
		\node[anchor=east] at (-0.55,-3) {$3$};

		
		\foreach \x in {0,1,2,3}{
			\foreach \y in {0,-1,-2,-3}{
				\fill (\x,\y) circle (0.65pt);
			}
		}

		
		\draw[line width=0.8pt]
		(1,0)
		--
		(0,-1)
		--
		(1,-2)
		--
		(2,-1)
		--
		cycle;

		
		\draw[line width=0.8pt]
		(2,-1)
		--
		(1,-2)
		--
		(2,-3)
		--
		(3,-2)
		--
		cycle;

		
		\foreach \p in {
			(1,0),
			(0,-1),
			(2,-1),
			(1,-2),
			(3,-2),
			(2,-3)
		}{
			\fill \p circle (1.4pt);
		}

		
		\node[anchor=west] at (4.55,-0.25)
		{\large $\Rightarrow\ \chi_q\!\left(V(Y_{1,0})\right)=$};
		
		\node[anchor=west] at (5.35,-1.55)
		{\large $Y_{1,0}+Y_{1,2}^{-1}Y_{2,1}+Y_{2,3}^{-1}$};
		
		\node[anchor=west] at (4.55,-2.75)
		{\large $\Rightarrow\ 3\ \mathrm{dim.}$};
		
	\end{tikzpicture}
	
	\caption{Path representation of $V(Y_{1,0})$ with q-character given in \eqref{qch10} and total dimension $3$.}
	\label{fig:fund_qChar}
	
\end{figure}
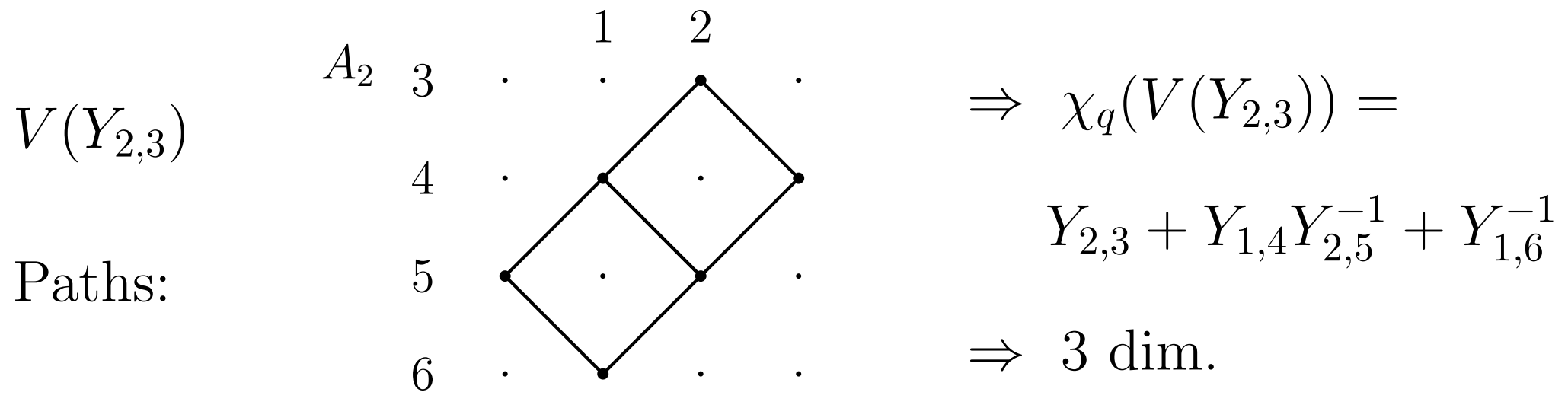
\begin{figure}[htbp]
	\centering
	
	\begin{tikzpicture}[
		x=0.85cm,
		y=0.85cm,
		baseline=(current bounding box.center)
		]
		
		
		\node[anchor=west] at (-5.2,-0.55) {\large $V(Y_{2,3})$};
		\node[anchor=west] at (-5.2,-2.05) {\large Paths:};

		
		\node[anchor=east] at (-1.15,0.15) {$A_2$};
		
		\node at (1,0.55) {$1$};
		\node at (2,0.55) {$2$};
		
		\node[anchor=east] at (-0.55, 0) {$3$};
		\node[anchor=east] at (-0.55,-1) {$4$};
		\node[anchor=east] at (-0.55,-2) {$5$};
		\node[anchor=east] at (-0.55,-3) {$6$};

		
		\foreach \x in {0,1,2,3}{
			\foreach \y in {0,-1,-2,-3}{
				\fill (\x,\y) circle (0.65pt);
			}
		}

		
		\draw[line width=0.8pt]
		(2,0)
		--
		(1,-1)
		--
		(2,-2)
		--
		(3,-1)
		--
		cycle;

		
		\draw[line width=0.8pt]
		(1,-1)
		--
		(0,-2)
		--
		(1,-3)
		--
		(2,-2)
		--
		cycle;

		
		\foreach \p in {
			(2,0),
			(1,-1),
			(3,-1),
			(0,-2),
			(2,-2),
			(1,-3)
		}{
			\fill \p circle (1.4pt);
		}

		
		\node[anchor=west] at (4.55,-0.25)
		{\large $\Rightarrow\ \chi_q\!\left(V(Y_{2,3})\right)=$};
		
		\node[anchor=west] at (5.35,-1.55)
		{\large $Y_{2,3}+Y_{1,4}Y_{2,5}^{-1}+Y_{1,6}^{-1}$};
		
		\node[anchor=west] at (4.55,-2.75)
		{\large $\Rightarrow\ 3\ \mathrm{dim.}$};
		
	\end{tikzpicture}
	
	\caption{Path representation of $V(Y_{2,3})$ with q-character given in \eqref{qch23} and total dimension $3$.}
	\label{fig:antifund_qChar}
	
\end{figure}
Here, we chose the spectral parameters such that both representations are in special position with respect to each other, i.e. $0$ and $3$.\footnote{ Any two loop parameters $a$ and $a\pm 3$ are in special position here \cite{C2002}.} Multiplying them together, we obtain the $q$-character of the adjoint representation and the $q$-character of the trivial representation. Thus, the $q$-character of the adjoint representation is given by
\begin{equation}\label{qchar8}
	\chi_q(V(Y_{1,0}Y_{2,3})) =(Y_{1,0}+Y_{1,2}^{-1}Y_{2,1}+Y_{2,3}^{-1})(Y_{2,3}+Y_{1,4}Y_{2,5}^{-1}+Y_{1,6}^{-1})-1.
\end{equation}
It can be obtained through the path formula as shown in Figure~\ref{fig:adj_qChar}.
If we now fuse another fundamental representation, there are two possible choices: $V(Y_{1,-2})$ or $V(Y_{1,6})$. Their respective $q$-characters are
\begin{align}
	\chi_q(V(Y_{1,-2})) &= Y_{1,-2}+Y_{1,0}^{-1}Y_{2,-1}+Y_{2,1}^{-1}, \text{ and}\\
	\chi_q(V(Y_{1,6})) &= Y_{1,6}+Y_{1,8}^{-1}Y_{2,7}+Y_{2,9}^{-1}.
\end{align}
Multiplying each of these by the $q$-character of the adjoint representation and checking the dominant monomials, we obtain the two irreducible composition factors corresponding to the dominant monomials $Y_{1,-2}Y_{1,0}Y_{2,3}$ and $Y_{2,-1}Y_{2,3}$ in the first case, and $Y_{1,0}Y_{2,3}Y_{1,6}$ and $Y_{1,0}$ in the second case. Figures~\ref{fig:tensoring_L(Y_{1,-2})} and~\ref{fig:tensoring_L(Y_{1,6})} visualise our findings by drawing corresponding snakes and Young diagrams.\footnote{Equations in the Grothendieck ring, to be precise.}
\begin{figure}[htbp]
	\centering
	
	\begin{tikzpicture}[
		x=0.62cm,
		y=0.62cm,
		line cap=round,
		line join=round
		]
		
		
		
		\begin{scope}[shift={(0,0)}]
			
			\node at (0,0.85) {$1$};
			\node at (1.25,0.85) {$2$};
			
			\node[anchor=east] at (-1.10, 0.00) {$-2$};
			\node[anchor=east] at (-1.10,-1.20) {$0$};
			\node[anchor=east] at (-1.10,-2.40) {$2$};
			\node[anchor=east] at (-1.10,-3.60) {$4$};
			
			\foreach \y in {0,-1.2,-2.4,-3.6}{
				\fill (0,\y) circle (0.50pt);
			}
			
			\foreach \y in {-0.6,-1.8,-3.0}{
				\fill (1.25,\y) circle (0.50pt);
			}
			
			\fill (0,0) circle (1.45pt);
			
		\end{scope}

		
		\begin{scope}[shift={(4.0,0)}]
			
			\node at (0,0.85) {$1$};
			\node at (1.25,0.85) {$2$};
			
			\foreach \y in {0,-1.2,-2.4,-3.6}{
				\fill (0,\y) circle (0.50pt);
			}
			
			\foreach \y in {-0.6,-1.8,-3.0}{
				\fill (1.25,\y) circle (0.50pt);
			}
			
			\fill (0,-1.20) circle (1.45pt);
			\fill (1.25,-3.00) circle (1.45pt);
			
			\draw[line width=0.78pt]
			(0,-1.20)--(1.25,-3.00);
			
		\end{scope}

		
		\node[font=\large] at (7.15,-1.75) {$=$};

		
		\begin{scope}[shift={(9.0,0)}]
			
			\node at (0,0.85) {$1$};
			\node at (1.25,0.85) {$2$};
			
			\foreach \y in {0,-1.2,-2.4,-3.6}{
				\fill (0,\y) circle (0.50pt);
			}
			
			\foreach \y in {-0.6,-1.8,-3.0}{
				\fill (1.25,\y) circle (0.50pt);
			}
			
			\fill (0,0) circle (1.45pt);
			\fill (0,-1.20) circle (1.45pt);
			\fill (1.25,-3.00) circle (1.45pt);
			
			\draw[line width=0.78pt]
			(0,0)--(0,-1.20)--(1.25,-3.00);
			
		\end{scope}

		
		\node[font=\large] at (12.55,-1.75) {$+$};

		
		\begin{scope}[shift={(14.1,0)}]
			
			\node at (0,0.85) {$1$};
			\node at (1.25,0.85) {$2$};
			
			\foreach \y in {0,-1.2,-2.4,-3.6}{
				\fill (0,\y) circle (0.50pt);
			}
			
			\foreach \y in {-0.6,-1.8,-3.0}{
				\fill (1.25,\y) circle (0.50pt);
			}
			
			\fill (1.25,-0.60) circle (1.45pt);
			
		\end{scope}

		
		\node[font=\large] at (16.95,-1.75) {$\cdot$};

		
		\begin{scope}[shift={(18.1,0)}]
			
			\node at (0,0.85) {$1$};
			\node at (1.25,0.85) {$2$};
			
			\foreach \y in {0,-1.2,-2.4,-3.6}{
				\fill (0,\y) circle (0.50pt);
			}
			
			\foreach \y in {-0.6,-1.8,-3.0}{
				\fill (1.25,\y) circle (0.50pt);
			}
			
			\fill (1.25,-3.00) circle (1.45pt);
			
		\end{scope}

		
		\node[
		align=center,
		anchor=north,
		font=\small
		] at (15.15,-4.35)
		{
			Yangian irreducible\\
			because the loop parameters\\
			are not in special position.
		};

		
		\node[anchor=west,font=\normalsize] at (-1.4,-6.85)
		{In Young diagrams:};

		
		\begin{scope}[shift={(0.0,-8.00)},scale=0.70]
			\draw (0,0) rectangle (1,1);
		\end{scope}
		
		\node at (0.35,-9.10) {$3$};

		\node[font=\large] at (2.05,-7.70) {$\cdot$};

		
		\begin{scope}[shift={(3.35,-8.00)},scale=0.70]
			\draw (0,0) rectangle (1,1);
			\draw (1,0) rectangle (2,1);
			\draw (0,-1) rectangle (1,0);
		\end{scope}
		
		\node at (4.05,-9.10) {$8$};

		\node[font=\large] at (6.25,-7.70) {$=$};

		
		\begin{scope}[shift={(7.65,-8.00)},scale=0.70]
			\draw (0,0) rectangle (1,1);
			\draw (1,0) rectangle (2,1);
			\draw (2,0) rectangle (3,1);
			\draw (0,-1) rectangle (1,0);
		\end{scope}
		
		\node at (8.70,-9.10) {$15$};

		\node[font=\large] at (11.35,-7.70) {$+$};

		
		\begin{scope}[shift={(13.00,-8.00)},scale=0.70]
			\draw (0,0) rectangle (1,1);
			\draw (0,-1) rectangle (1,0);
		\end{scope}
		
		\node at (13.35,-9.10) {$\overline{3}$};

		\node[font=\large] at (15.15,-7.70) {$\cdot$};

		
		\begin{scope}[shift={(16.70,-8.00)},scale=0.70]
			\draw (0,0) rectangle (1,1);
			\draw (0,-1) rectangle (1,0);
		\end{scope}
		
		\node at (17.05,-9.10) {$\overline{3}$};

		
		\draw[
		decorate,
		decoration={
			brace,
			amplitude=5pt
		},
		line width=0.65pt
		]
		(12.65,-7.05)
		--
		(17.65,-7.05);

	\end{tikzpicture}
	
	\caption{
		Grothendieck-ring identity obtained by tensoring with $V(Y_{1,-2})$.
	}
	\label{fig:tensoring_L(Y_{1,-2})}
	
\end{figure}
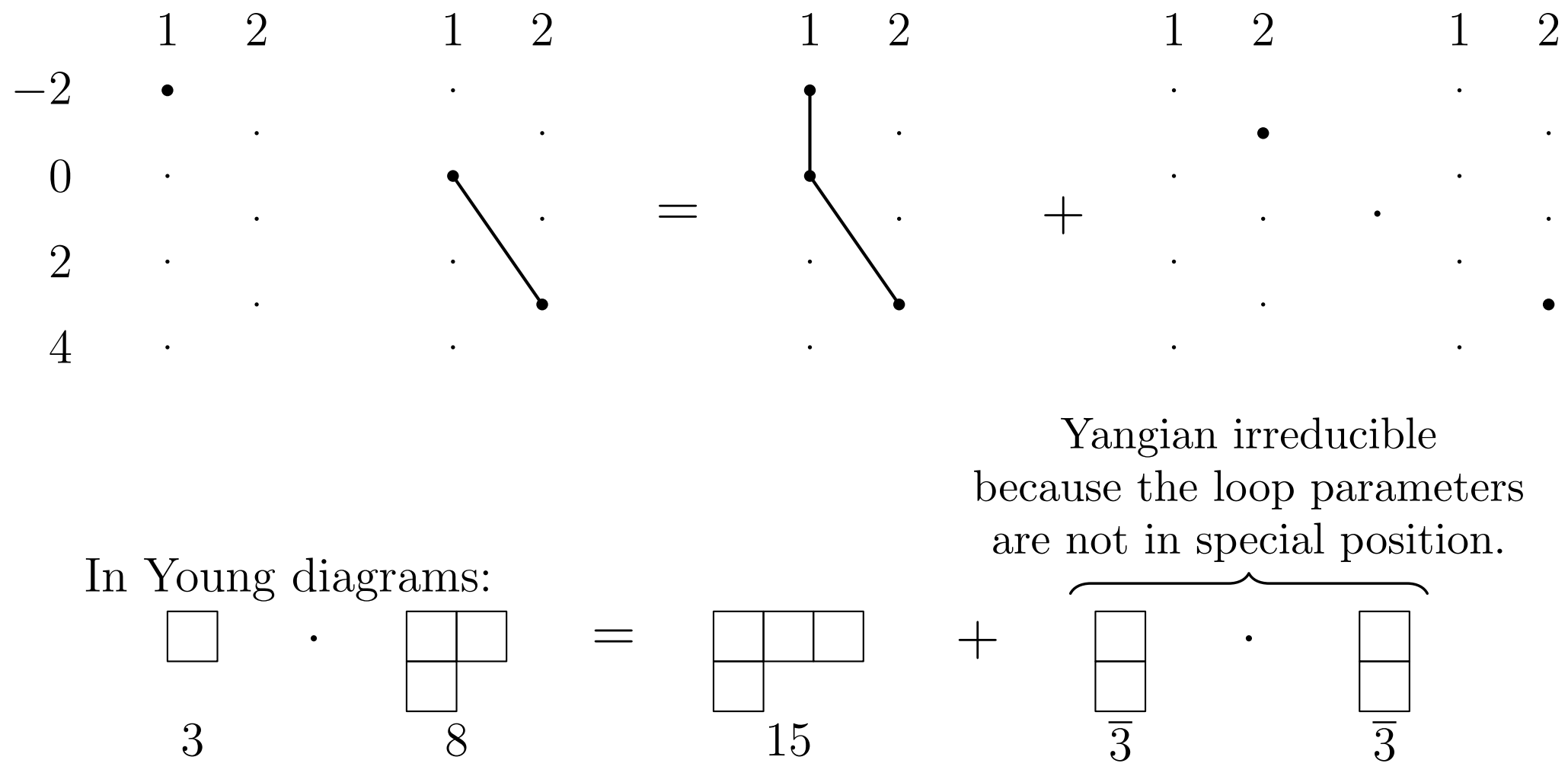
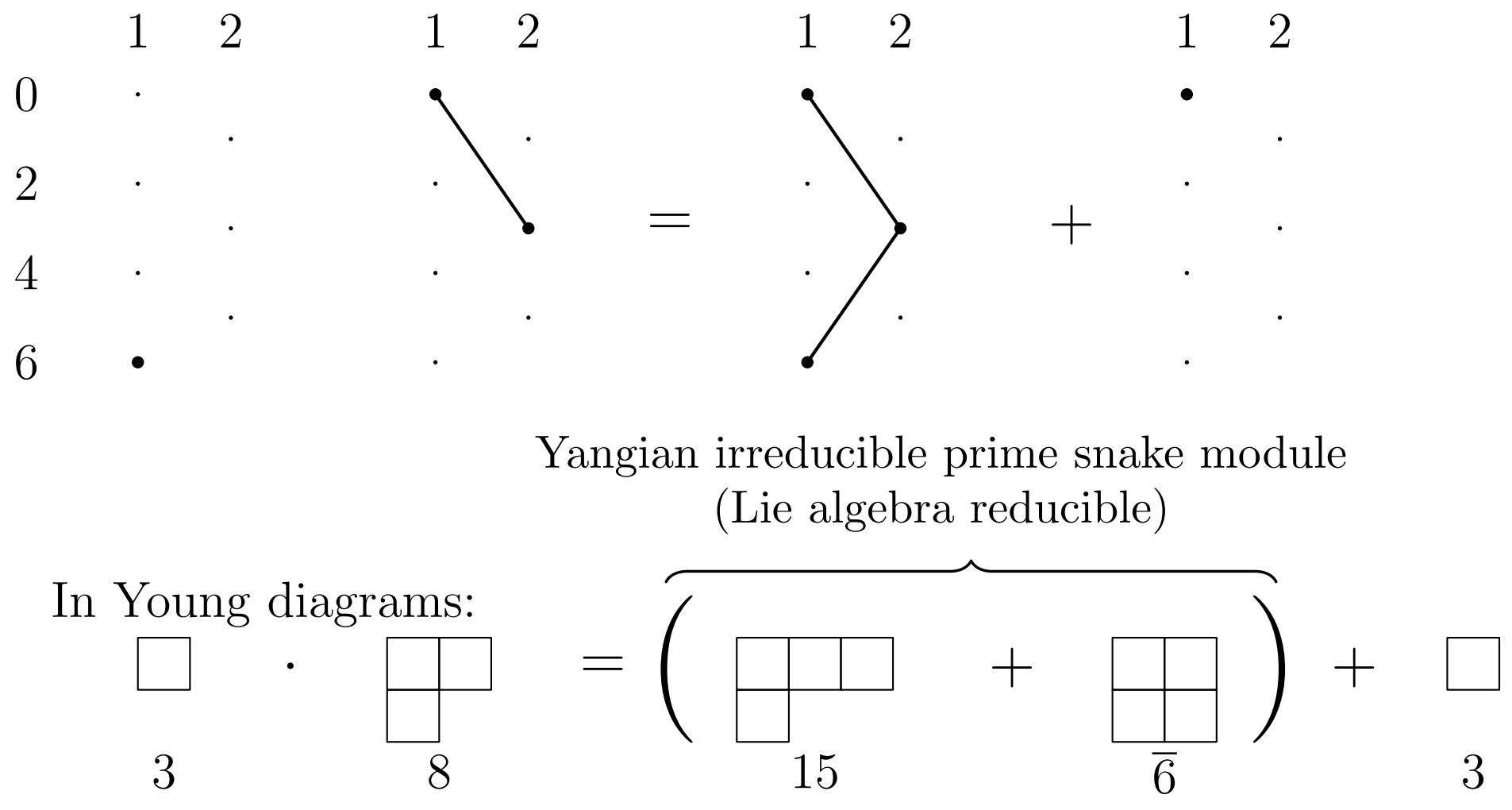
\begin{figure}[htbp]
	\centering
	
	\begin{tikzpicture}[
		x=0.62cm,
		y=0.62cm,
		line cap=round,
		line join=round
		]
		

		
		\begin{scope}[shift={(0,0)}]
			
			\node at (0,0.85) {$1$};
			\node at (1.25,0.85) {$2$};
			
			\node[anchor=east] at (-1.10, 0.00) {$0$};
			\node[anchor=east] at (-1.10,-1.20) {$2$};
			\node[anchor=east] at (-1.10,-2.40) {$4$};
			\node[anchor=east] at (-1.10,-3.60) {$6$};
			
			\foreach \y in {0,-1.2,-2.4,-3.6}{
				\fill (0,\y) circle (0.50pt);
			}
			
			\foreach \y in {-0.6,-1.8,-3.0}{
				\fill (1.25,\y) circle (0.50pt);
			}
			
			\fill (0,-3.60) circle (1.45pt);
			
		\end{scope}

		
		\begin{scope}[shift={(4.0,0)}]
			
			\node at (0,0.85) {$1$};
			\node at (1.25,0.85) {$2$};
			
			\foreach \y in {0,-1.2,-2.4,-3.6}{
				\fill (0,\y) circle (0.50pt);
			}
			
			\foreach \y in {-0.6,-1.8,-3.0}{
				\fill (1.25,\y) circle (0.50pt);
			}
			
			\fill (0,0) circle (1.45pt);
			\fill (1.25,-1.80) circle (1.45pt);
			
			\draw[line width=0.78pt]
			(0,0)--(1.25,-1.80);
			
		\end{scope}

		
		\node[font=\large] at (7.15,-1.75) {$=$};

		%
		
		\begin{scope}[shift={(9.0,0)}]
			
			\node at (0,0.85) {$1$};
			\node at (1.25,0.85) {$2$};
			
			\foreach \y in {0,-1.2,-2.4,-3.6}{
				\fill (0,\y) circle (0.50pt);
			}
			
			\foreach \y in {-0.6,-1.8,-3.0}{
				\fill (1.25,\y) circle (0.50pt);
			}
			
			\fill (0,0) circle (1.45pt);
			\fill (1.25,-1.80) circle (1.45pt);
			\fill (0,-3.60) circle (1.45pt);
			
			\draw[line width=0.78pt]
			(0,0)
			--
			(1.25,-1.80)
			--
			(0,-3.60);
			
		\end{scope}

		
		\node[font=\large] at (12.55,-1.75) {$+$};

		
		\begin{scope}[shift={(14.1,0)}]
			
			\node at (0,0.85) {$1$};
			\node at (1.25,0.85) {$2$};
			
			\foreach \y in {0,-1.2,-2.4,-3.6}{
				\fill (0,\y) circle (0.50pt);
			}
			
			\foreach \y in {-0.6,-1.8,-3.0}{
				\fill (1.25,\y) circle (0.50pt);
			}
			
			\fill (0,0) circle (1.45pt);
			
		\end{scope}

		
		\node[
		align=center,
		anchor=north,
		font=\small
		] at (10.80,-4.35)
		{
			Yangian irreducible prime snake module \\(Lie algebra
			reducible)
		};

		
		\node[anchor=west,font=\normalsize] at (-1.4,-6.85)
		{In Young diagrams:};

		
		\begin{scope}[shift={(0.0,-8.00)},scale=0.70]
			
			\draw (0,0) rectangle (1,1);
			
		\end{scope}
		
		\node at (0.35,-9.10) {$3$};

		
		\node[font=\large] at (2.05,-7.70) {$\cdot$};

		
		\begin{scope}[shift={(3.35,-8.00)},scale=0.70]
			
			\draw (0,0) rectangle (1,1);
			\draw (1,0) rectangle (2,1);
			\draw (0,-1) rectangle (1,0);
			
		\end{scope}
		
		\node at (4.05,-9.10) {$8$};

		
		\node[font=\large] at (6.25,-7.70) {$=$};

		
		\node[
		font=\fontsize{35}{35}\selectfont
		] at (7.20,-7.72) {$($};

		
		\begin{scope}[shift={(8.05,-8.00)},scale=0.70]
			
			\draw (0,0) rectangle (1,1);
			\draw (1,0) rectangle (2,1);
			\draw (2,0) rectangle (3,1);
			\draw (0,-1) rectangle (1,0);
			
		\end{scope}
		
		\node at (9.10,-9.10) {$15$};

		
		\node[font=\large] at (11.75,-7.70) {$+$};

		%
		
		\begin{scope}[shift={(13.10,-8.00)},scale=0.70]
			
			\draw (0,0) rectangle (1,1);
			\draw (1,0) rectangle (2,1);
			\draw (0,-1) rectangle (1,0);
			\draw (1,-1) rectangle (2,0);
			
		\end{scope}
		
		\node at (13.80,-9.10) {$\overline{6}$};

		\node[
		font=\fontsize{35}{35}\selectfont
		] at (15.25,-7.72) {$)$};

		
		\node[font=\large] at (16.35,-7.70) {$+$};

		
		\begin{scope}[shift={(17.60,-8.00)},scale=0.70]
			
			\draw (0,0) rectangle (1,1);
			
		\end{scope}
		
		\node at (17.95,-9.10) {$3$};

		
		\draw[
		decorate,
		decoration={
			brace,
			amplitude=5pt
		},
		line width=0.65pt
		]
		(7.10,-6.55)
		--
		(15.30,-6.55);

	\end{tikzpicture}
	
	\caption{
		Grothendieck-ring identity obtained by tensoring with $V(Y_{1,6})$.
	}
	\label{fig:tensoring_L(Y_{1,6})}
	
\end{figure}
Hence, the two fusion procedures exhibit Yangian-irreducible representations which are reducible as $\mathfrak{gl}_3$-modules. In the first case, the relevant $9$-dimensional representation is not prime as a Yangian module. Its prime factors are $\mathfrak{gl}_3$-irreducible, and the corresponding Lax operator therefore factorises into Lax operators associated with Lie-algebra irreducibles.

The second case, by contrast, yields a $21$-dimensional prime snake module which is irreducible as a Yangian module but reducible as a $\mathfrak{gl}_3$-module. This is the first example in our discussion for which the corresponding prime irreducible Lax operator is neither linear nor obtained from the inversion construction. It will serve as our main non-trivial example below.

	


\subsection{The Gelfand--Tsetlin algebra and loop weights}
\label{subsec:GT_algebra}

We now construct explicit basis vectors of the skew representations
and determine their loop weights. Following \cite{NT1998}, we use
common eigenvectors of commuting operators obtained from quantum
minors of the Lax operator. For ordered sets
$I=(i_1,\ldots,i_m)$ and $J=(j_1,\ldots,j_m)$, we define
\begin{equation}
	Q_{I,J}(u)
	\coloneqq
	\sum_{\sigma\in S_m}\operatorname{sgn}(\sigma)
	L_{i_1,j_{\sigma(1)}}(u)
	L_{i_2,j_{\sigma(2)}}(u-1)\cdots
	L_{i_m,j_{\sigma(m)}}(u-m+1).
	\label{eqn:quantum_minor_GT}
\end{equation}
Compared with ordinary matrix minors, the spectral parameter is
shifted by one between consecutive factors. In particular, we use
\begin{align}
	A_m(u)&\coloneqq Q_{(1,\ldots,m),(1,\ldots,m)}(u),
	\label{eqn:GT_A_def}\\
	B_m(u)&\coloneqq Q_{(1,\ldots,m),(1,\ldots,m-1,m+1)}(u),
	\label{eqn:GT_B_def}\\
	C_m(u)&\coloneqq Q_{(1,\ldots,m-1,m+1),(1,\ldots,m)}(u),
	\label{eqn:GT_C_def}
\end{align}
where $A_m$ is defined for $m=1,\ldots,n$ and $B_m,C_m$ for
$m=1,\ldots,n-1$. The coefficients of the $A_m(u)$ commute
and generate the Gelfand--Tsetlin algebra; $A_n(u)=\detq L(u)$
is the quantum determinant. Their common eigenvectors form
the Gelfand--Tsetlin basis of the skew representation.
The operators $B_m(u)$ and $C_m(u)$ provide the ladder steps
between these vectors.

To determine the loop weights, we relate the $A_m(u)$ to the
diagonal entries $D_i(u)$ of the Gauss decomposition
$L(u)=L^-(u)D(u)L^+(u)$.
For our convention $R(u)=uI+\operatorname{P}$, we have
\begin{equation}
	A_m(u)=D_1(u-m+1)D_2(u-m+2)\cdots D_m(u).
	\label{eqn:GT_A_Gauss}
\end{equation}
This follows from \cite[Eq.~(8.5) and Theorem~8.6]{BrundanKleshchev2005}
by replacing its matrix $T(x)$ with $L(-x)$ and evaluating its
quantum minor at $x=-u+m-1$. Setting $A_0(u)\coloneqq1$, we obtain
\begin{equation}
	D_m(u)=\frac{A_m(u)}{A_{m-1}(u-1)}.
	\label{eqn:GT_D_from_A}
\end{equation}
Hence, for a common eigenvector $\xi_\Lambda$ with
\begin{equation}
	A_m(u)\xi_\Lambda=a_m^\Lambda(u)\xi_\Lambda,
	\qquad m=1,\ldots,n,
	\label{eqn:GT_A_eigenvalues}
\end{equation}
\eqref{eqn:Gauß_dec1} gives the loop-weight functions
\begin{equation}
	\psi_i^\Lambda(u)
	=
	\frac{a_{i+1}^\Lambda(u)\,a_{i-1}^\Lambda(u-1)}
	{a_i^\Lambda(u)\,a_i^\Lambda(u-1)},
	\qquad a_0^\Lambda(u)\coloneqq1.
	\label{eqn:GT_to_loop_weight}
\end{equation}

Under $L(u)\mapsto g(u)L(u)$, every quantum minor of size $m$
acquires the same scalar factor. In particular,
\begin{equation}
	A_m(u)\longmapsto
	\left(\prod_{j=0}^{m-1}g(u-j)\right)A_m(u).
	\label{eqn:GT_scalar_normalisation}
\end{equation}
The same factor multiplies $B_m(u)$ and $C_m(u)$.
This changes the quantum determinant $A_n(u)$, but leaves the
loop weights unchanged because the scalar factors cancel
in \eqref{eqn:GT_to_loop_weight}.

To see how $B_m(u)$ connects these eigenvectors, we use the relations
$[A_l(u),B_m(v)]=0$ for $l\neq m$ and
\begin{equation}
	(u-v)[A_m(u),B_m(v)]
	=
	B_m(u)A_m(v)-B_m(v)A_m(u)
	\label{eqn:GT_AB_relation}
\end{equation}
from \cite[Proposition~1.9]{NT1998}.
If $a_m^\Lambda(v)=0$ and $B_m(v)\xi_\Lambda\neq0$, then
$B_m(v)\xi_\Lambda$ is again a common eigenvector. All eigenvalues
remain unchanged except for
\begin{equation}
	a_m^\Lambda(u)
	\longmapsto
	\frac{u-v-1}{u-v}\,a_m^\Lambda(u).
	\label{eqn:GT_root_shift}
\end{equation}
Thus one zero moves from $v$ to $v+1$. Starting from the
lowest-weight vector, we repeat these steps with the updated
eigenvalues to construct further basis vectors.

\medskip

\paragraph{\textit{Normalisation of the ladder operators.}} \leavevmode\par
The scalar normalisation of $L(u)$ leaves the loop weights
unchanged, but it can affect the evaluation of the ladder
operators through the factors in
\eqref{eqn:GT_scalar_normalisation}. If such a factor
vanishes at a zero $v$ used for a ladder step, it increases the
multiplicity of this zero in $a_m^\Lambda(u)$ and makes
$B_m(v)\xi_\Lambda$ vanish. The corresponding step can therefore
be recovered by removing the common factor before evaluating
the operator. To work directly with the polynomial minors used
below, we compute this value by differentiation as follows.

Let $v$ be a zero of $a_m^\Lambda(u)$ of multiplicity $\mu$ and set
\begin{equation}
	F_m(z)\coloneqq B_m(z)\xi_\Lambda.
\end{equation}
Suppose $F_m(z)$ has a zero of order $r$, with $1\leq r<\mu$.
Then we can write
\begin{equation*}
	F_m(z)=(z-v)^r\widetilde F_m(z),
	\qquad \widetilde F_m(v)\neq0.
\end{equation*}
Removing the factor $(z-v)^r$ gives the non-zero vector
\begin{equation}
	\xi_{\Lambda'}
	\coloneqq
	\lim_{z\to v}\frac{B_m(z)\xi_\Lambda}{(z-v)^r}
	=
	\frac{1}{r!}
	\left.\frac{\mathrm d^r}{\mathrm dz^r}
	B_m(z)\xi_\Lambda\right|_{z=v}.
	\label{eqn:confluent_GT_vector}
\end{equation}
The equality follows by expanding $B_m(z)\xi_\Lambda$ around
$z=v$. Thus the derivative gives the value obtained after
removing the vanishing factor from $B_m(z)\xi_\Lambda$.

To see why this vector gives the same ladder step, apply
\eqref{eqn:GT_AB_relation} to $\xi_\Lambda$:
\begin{equation}
	(u-z)A_m(u)F_m(z)
	=
	(u-z-1)a_m^\Lambda(u)F_m(z)
	+
	a_m^\Lambda(z)B_m(u)\xi_\Lambda.
	\label{eqn:confluent_GT_proof}
\end{equation}
Divide by $(z-v)^r$ and let $z\to v$. Since $\mu>r$, the last
term still contains at least one factor of $z-v$ and vanishes.
The remaining terms give exactly the eigenvalue transformation
\eqref{eqn:GT_root_shift}. For $l\neq m$, the relation
$[A_l(u),B_m(z)]=0$ ensures that the eigenvalues
remain unchanged.

If $A_m(u)$ and $B_m(u)$ share a scalar polynomial factor $h(u)$
on the module, we can remove it before constructing the ladder:
\begin{equation*}
	A_m(u)=h(u)\widehat A_m(u),
	\qquad B_m(u)=h(u)\widehat B_m(u).
\end{equation*}
Cancelling $h(u)h(v)$ shows that the normalised operators satisfy
the same relation \eqref{eqn:GT_AB_relation}. At a simple zero
$v$ of $h$, the product rule gives
$B_m'(v)=h'(v)\widehat B_m(v)$, so a derivative step is an
ordinary evaluation of $\widehat B_m$ up to a non-zero scalar.
This is what happens in the $21$-dimensional example below.
    
\section{Snake/Skew Lax operators, their eigenvalues and the oscillator construction}
	\label{sect:snake/skew_lax}
	
	We determine the lowest-weight eigenvalues of snake/skew Lax operators
	and construct their oscillator realisations. We begin with the
	$\mathfrak{gl}_3$ case and then decompose general $\mathfrak{gl}_n$
	skew diagrams into shifted Young diagrams. These blocks determine
	the Drinfeld polynomials and provide the evaluation modules used
	in the oscillator construction below.
	
	\subsection{Eigenvalues of the \texorpdfstring{$\mathfrak{gl}_3$}{gl₃} snake/skew Lax on the lowest weight}
	We first consider the following class of $\mathfrak{gl}_3$ skew
	diagrams $\lambda\backslash\mu$, with
	\begin{equation}
		\lambda = (\lambda_1,\lambda_2,\dots,\lambda_k,\lambda_{k+1}),\quad \mu = (\mu_1,\mu_2,\dots,\mu_{k-2})\label{eqn:gl_3_skew_diag}
	\end{equation}
	and introduce the auxiliary endpoint values
	\begin{equation}
		\mu_{k-1}\coloneqq \lambda_{k+1},\quad \mu_0 \coloneqq \lambda_1.
	\end{equation}
	These values are not additional parts of the partition $\mu$.
	We take $k\geq2$ and assume $\mu_{k-1}\leq\lambda_k$,
	$\lambda_2\leq\mu_0$, and
	$\lambda_{r+1}<\mu_{r-1}<\lambda_r$ for $r=2,\ldots,k-1$.
	We exclude diagrams consisting entirely of full columns,
	which give one-dimensional representations.
	Full columns of height three are allowed only at the left edge.
	Remark~\ref{rem:full_columns} below explains this choice in view of
	primality and the scalar factors contributed by full columns
	at the right edge.
	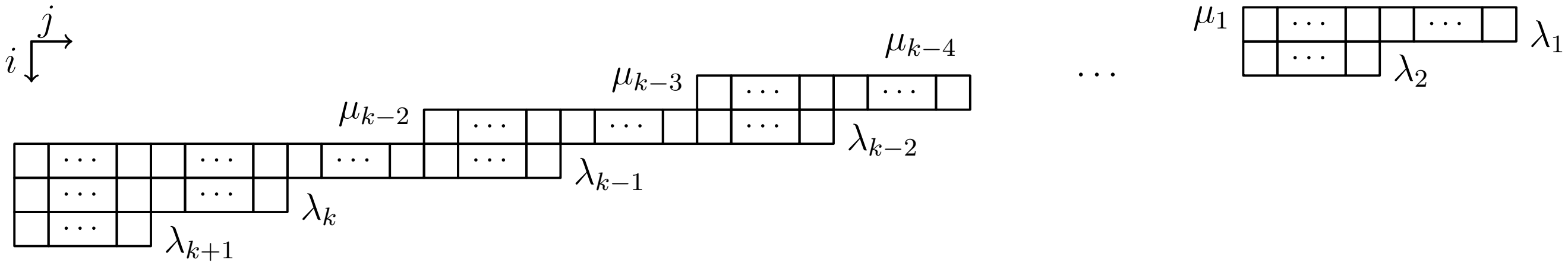
\begin{figure}[htbp]
		\centering
		
		\begin{tikzpicture}[
			x=0.38cm,
			y=0.38cm,
			line cap=round,
			line join=round
			]
			
			
			\newcommand{\skewblock}[2]{%
				\draw[line width=0.72pt]
				(#1,#2) rectangle ++(1,1);
				
				\draw[line width=0.72pt]
				({#1+1},#2) rectangle ++(2,1);
				
				\node[scale=0.82]
				at ({#1+2},{#2+0.50})
				{$\cdots$};
				
				\draw[line width=0.72pt]
				({#1+3},#2) rectangle ++(1,1);
			}

			
			\coordinate (axisorigin) at (0.50,6.00);
			
			\draw[->,line width=0.75pt]
			(axisorigin)
			--
			++(1.20,0);
			
			\node at (0.95,6.58) {$j$};
			
			\draw[->,line width=0.75pt]
			(axisorigin)
			--
			++(0,-1.20);
			
			\node at (-0.10,5.45) {$i$};

			
			\skewblock{0}{0}
			
			\skewblock{0}{1}
			\skewblock{4}{1}
			
			\skewblock{0}{2}
			\skewblock{4}{2}
			\skewblock{8}{2}
			\skewblock{12}{2}
			
			\skewblock{12}{3}
			\skewblock{16}{3}
			\skewblock{20}{3}
			
			\skewblock{20}{4}
			\skewblock{24}{4}

			%
			
			\node[
			anchor=west
			] at (4.00,0.08)
			{$\lambda_{k+1}$};
			
			\node[
			anchor=west
			] at (8.00,1.08)
			{$\lambda_k$};
			
			\node[
			anchor=west
			] at (16.00,2.08)
			{$\lambda_{k-1}$};
			
			\node[
			anchor=west
			] at (24.00,3.08)
			{$\lambda_{k-2}$};

			
			\node[
			anchor=south east
			] at (12.00,3.12)
			{$\mu_{k-2}$};
			
			\node[
			anchor=south east
			] at (20.00,4.12)
			{$\mu_{k-3}$};
			
			\node[
			anchor=south east
			] at (28.00,5.12)
			{$\mu_{k-4}$};

			
			\node[
			scale=0.95
			] at (31.8,5.00)
			{$\cdots$};

			
			\skewblock{36}{5}
			
			\skewblock{36}{6}
			\skewblock{40}{6}

			%
			
			\node[
			anchor=south east
			] at (36.00,5.95)
			{$\mu_1$};
			
			\node[
			anchor=west
			] at (40.00,5.15)
			{$\lambda_2$};
			
			\node[
			anchor=west
			] at (44.00,6.15)
			{$\lambda_1$};

		\end{tikzpicture}
		
		\caption{
			The class of $\mathfrak{gl}_3$ skew diagrams considered here,
			with full columns of height three allowed at the left edge.
		}
		\label{fig:skew_sl3}
		
	\end{figure}
	The columns of height one and two determine the corresponding prime snake.
	We use the row and column coordinates introduced in \eqref{eqn:snake_skew}. Recall that a column of height $i$
	whose bottom box has coordinates $(s,t)$ corresponds to the snake point
	\[
	(i,k_t),\qquad k_t=2(t-s)+i-1.
	\]
	For the columns of height two in Figure~\ref{fig:skew_sl3}, the bottom box
	lies in row $s=k-l+1$. Hence
	\begin{equation}
		(i_t,k_t)
		=
		\bigl(2,\,2t-2(k-l+1)+1\bigr),
		\qquad
		\mu_{k-l}<t\leq\lambda_{k-l+1},
		\quad l=1,\dots,k-1.
	\end{equation}
	Similarly, for the columns of height one the bottom box lies in row
	$s=k-l$, and therefore
	\begin{equation}
		(i_t,k_t)
		=
		\bigl(1,\,2t-2(k-l)\bigr),
		\qquad
		\lambda_{k-l+1}<t\leq\mu_{k-l-1},
		\quad l=1,\dots,k-1.
	\end{equation}
	Thus the snake is read directly from the original coordinates of the skew
	diagram. We can now either use \eqref{eqn:DrinPoly_snake} or read the
	Drinfeld polynomials directly from the bottom-box contents as in
	\cite{NT1998}. We obtain
	\begin{align}
		\frac{P_1(u-1)}{P_1(u)}&= \frac{(u+\lambda_{k}-(k-1))(u+\lambda_{k-1}-(k-2))\cdots(u+\lambda_2-1)}{(u+\mu_{k-2}-(k-1))(u+\mu_{k-3}-(k-2))\cdots(u+\mu_0-1)}\quad \text{and}\\
		\frac{P_2(u-1)}{P_2(u)} &= \frac{(u+\mu_{k-1}-k)(u+\mu_{k-2}-(k-1))\cdots(u+\mu_1-2)}{(u+\lambda_{k}-k)(u+\lambda_{k-1}-(k-1))\cdots(u+\lambda_2-2)}.
	\end{align}
	By \eqref{eqn:Shifted_L_ii_to_Drin_polys_direct} and assuming that the eigenvalues of $L_{i,i}$ on $\xi$ are polynomials of GCD $1$, we finally obtain
	\begin{align}
		L_{11}(u)\xi =& (u+\mu_{k-2}-(k-2))(u+\mu_{k-3}-(k-3))\cdots(u+\mu_0)\xi\notag\\
		L_{22}(u)\xi =& (u+\lambda_{k}-(k-2))(u+\lambda_{k-1}-(k-3))\cdots(u+\lambda_2)\xi\notag\\
		L_{33}(u)\xi =& (u+\mu_{k-1}-(k-2))(u+\mu_{k-2}-(k-3))\cdots(u+\mu_1)\xi.\label{eqn:gl_3_lax_lowest_EV}
	\end{align}

	Corollary~\ref{cor:Osc_snake_Lax_sl3} realises these eigenvalues
	by restricting a product of $k-1$ monic linear oscillator Lax
	operators, so the resulting Lax matrix has degree $k-1$ in $u$.
	
	\medskip
\paragraph{\textit{Dimensions from the dominant loop weight.}} \leavevmode\par
The path formula also gives a dimension determinant. It is convenient
to state it for arbitrary $\mathfrak{gl}_n$, using the snake coordinates
introduced in Section~\ref{sect:gen_def}.

\begin{cor}[Dimension determinant]\label{cor:LGV_Det_dim}
Let $m=\prod_{a=1}^{T}Y_{i_a,k_a}$ be a snake monomial in type
$A_{n-1}$, with its factors in snake order. Then
\begin{equation}
    \dim V(m)=
    \det_{1\leq a,b\leq T}
    \left[\binom{n}{\frac{i_a+i_b+k_a-k_b}{2}}\right].
    \label{eqn:general_snake_dimension}
\end{equation}
Here $\binom{n}{r}=0$ if $r\notin\{0,1,\ldots,n\}$, and the
empty determinant is $1$. The same formula gives the dimension
of a skew $Y(\mathfrak{gl}_n)$-module with dominant
$Y(\mathfrak{sl}_n)$ loop weight $m$. $\odot$
\end{cor}

\begin{proof}
The initial point of the $a$-th path is $(0,i_a+k_a)$ and the
final point of the $b$-th path is $(n,n-i_b+k_b)$. A path between
these points has $n$ steps, of which
$(i_a+i_b+k_a-k_b)/2$ point downwards. Their number is therefore
the binomial coefficient in \eqref{eqn:general_snake_dimension}.
The snake condition strictly orders both the initial and final
heights. Since all paths have the same height parity at a fixed
horizontal coordinate, two paths cannot exchange their order
without sharing a vertex. Thus a non-intersecting family can
only connect the endpoints in the prescribed order.
The Lindstr\"om--Gessel--Viennot lemma \cite{GesselViennot1985}
identifies the determinant with the number of these families.
By Theorem~\ref{thm:path_formula} and thinness, this number is
$\dim V(m)$. One-dimensional determinant factors do not change
the dimension.
\end{proof}

This is the dual Jacobi--Trudi formula evaluated at $n$ unit
variables, written in snake coordinates. Indeed, let $\lambda/\mu$
be the diagram associated with $m$ by \eqref{eqn:snake_skew},
translated into positive rows, and let $\lambda',\mu'$ denote
the conjugate partitions. The column coordinates give
\begin{equation*}
    \frac{i_a+i_b+k_a-k_b}{2}
    =\lambda'_b-\mu'_a+a-b.
\end{equation*}
Transposing the matrix in \eqref{eqn:general_snake_dimension}
therefore yields
\begin{equation*}
    \dim V(m)=
    \det_{1\leq a,b\leq T}
    \left[\binom{n}{\lambda'_a-\mu'_b-a+b}\right]
    =s_{\lambda/\mu}(1^n),
\end{equation*}
where $s_{\lambda/\mu}$ is the skew Schur polynomial. The last
equality follows from \cite[Eq.~(42)]{vanLeeuwen2006} and
$e_r(1^n)=\binom{n}{r}$ for the elementary symmetric polynomials.

For the $\mathfrak{gl}_3$ diagrams above, set
\begin{equation*}
    r_{2s-1}=\mu_{s-1}-\lambda_{s+1},\qquad
    r_{2s}=\lambda_{s+1}-\mu_s,\qquad s=1,\ldots,k-1.
\end{equation*}
The determinant then has alternating diagonal blocks of sizes
$r_{2s-1}$ and $r_{2s}$. Their entries are
$\binom{3}{1+i-j}$ and $\binom{3}{1+j-i}$, respectively.
The entries connecting consecutive non-empty blocks are $1$
on either side of the diagonal. Figure~\ref{fig:dim_skew_sl3}
in Appendix~\ref{app:skew_dimension_determinant} displays this
specialisation. The general formula avoids constructing the
block matrix separately for each diagram.

	\subsection{The general \texorpdfstring{$\mathfrak{gl}_n$}{glₙ} snake/skew Lax}
	\label{subsec:general_skew_Lax}
	
	Consider a skew diagram $\omega=\lambda\backslash\mu$ with at most
	$n$ boxes in each column. We decompose it into disjoint blocks whose
	columns start in the same row. Each block is a shifted ordinary Young
	diagram. This will allow us both to read off the Drinfeld polynomials
	and to construct the Lax operator from evaluation representations.
	
	Here we extend the partitions by zeros and set
	$\mu_0\coloneqq\lambda_1$. For each $k\geq0$, take the boxes of
	$\omega$ in the columns
	\begin{equation}
		\mu_{k+1}<j\leq\mu_k.
		\label{eqn:gen_segm_ineq}
	\end{equation}
	Every non-empty column in this interval starts in row $k+1$.
	Its height weakly decreases as we move to the right, since the row
	lengths of $\lambda$ weakly decrease. Thus the boxes in the interval
	form an ordinary Young diagram shifted down by $k$ rows and right
	by $\mu_{k+1}$ columns. Empty blocks are omitted; in particular,
	$\mu_k=\mu_{k+1}$ contributes no block. The intervals partition all
	columns, so every box of $\omega$ belongs to exactly one block.
	
\begin{figure}[htbp]
		\centering
		\begin{tikzpicture}[x=0.65cm,y=-0.65cm,
			line cap=round,line join=round]
			\foreach \j in {1,2} {
				\draw[fill=black!5,draw=black!30,dashed]
				({\j-1},0) rectangle (\j,1);
			}
			\foreach \i in {2,3,4} {
				\foreach \j in {1,2} {
					\draw[fill=black!23,line width=0.5pt]
					({\j-1},{\i-1}) rectangle (\j,\i);
				}
			}
			\foreach \i/\last in {1/5,2/4,3/3} {
				\foreach \j in {3,...,\last} {
					\draw[fill=black!8,line width=0.5pt]
					({\j-1},{\i-1}) rectangle (\j,\i);
				}
			}
			\node at (1,0.5) {$\mu$};
			\draw[->] (-0.6,-0.6) -- (0.5,-0.6)
			node[right] {$j$};
			\draw[->] (-0.6,-0.6) -- (-0.6,0.5)
			node[below] {$i$};
			\node[anchor=north] at (1,4.2) {$B_2$};
			\node[anchor=south] at (3.5,-0.15) {$B_1$};
			\node[anchor=west,align=left] at (6.2,1.2)
			{$\nu^{(1)}=(3,2,1,0)$,\quad $s_1=2$\\[5pt]
				$\nu^{(2)}=(2,2,2,0)$,\quad $s_2=-1$};
		\end{tikzpicture}
		\caption{The disjoint Young blocks of
			$\lambda=(5,4,3,2)$ and $\mu=(2)$ for $n=4$.
			The right block $B_1$ contains columns $3,4,5$ and starts in
			row $1$; the left block $B_2$ contains columns $1,2$ and starts
			in row $2$. The dashed boxes belong to the removed diagram $\mu$.}
		\label{fig:skew_sln}
	\end{figure}
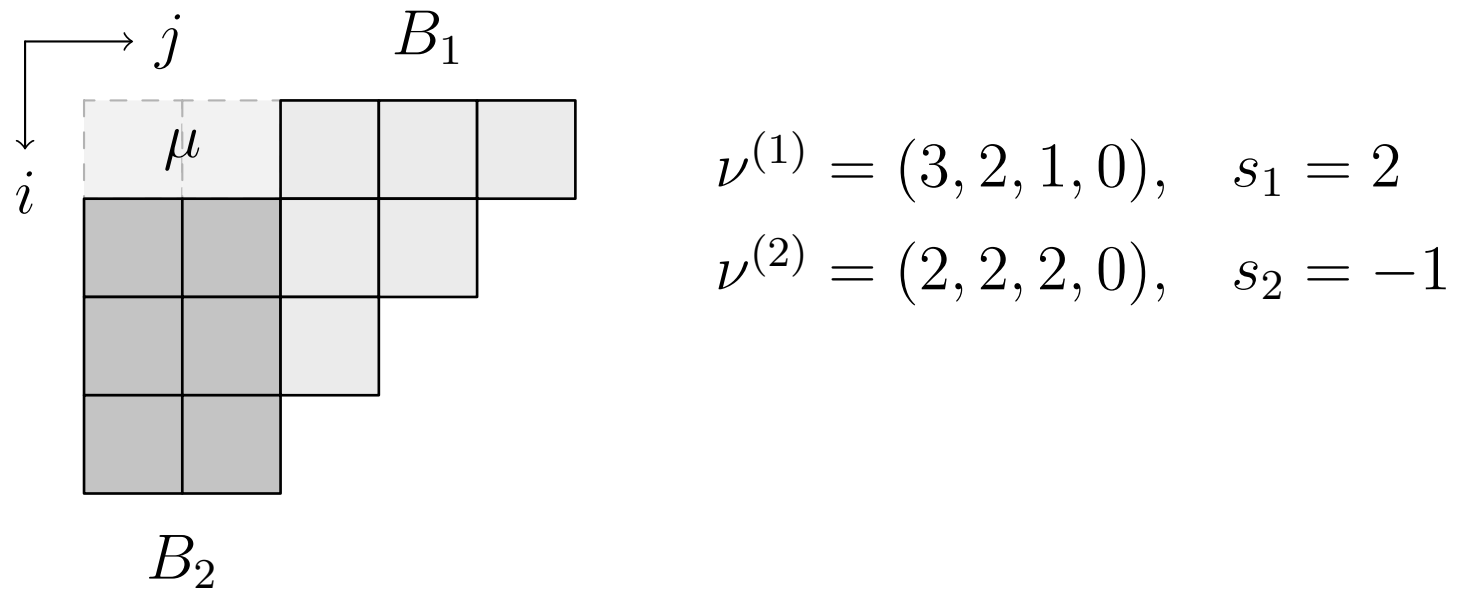

	Number the non-empty blocks $B_1,\ldots,B_K$ from right to left,
	and let $q_r$ be the index $k$ of $B_r$ in
	\eqref{eqn:gen_segm_ineq}. Thus $q_1<\cdots<q_K$.
	Put $p_r=\mu_{q_r+1}$ and denote the ordinary partition shape of
	$B_r$ by $\nu^{(r)}$, extended by zeros to length $n$. Explicitly,
	\begin{equation}
		\nu_i^{(r)}
		=\max\{0,\min(\lambda_{q_r+i},\mu_{q_r})-p_r\},
		\qquad i=1,\ldots,n,
		\label{eqn:skew_block_shape}
	\end{equation}
	and
	\begin{equation*}
		B_r=\{(q_r+i,p_r+j):1\leq i\leq n,
		\ 1\leq j\leq\nu_i^{(r)}\}.
	\end{equation*}
	The content of such a box is
	\begin{equation}
		(p_r+j)-(q_r+i)=j-i+s_r,
		\qquad s_r\coloneqq p_r-q_r.
		\label{eqn:skew_block_content_shift}
	\end{equation}
	The shift $s_r$ records the position of the block in the original
	skew diagram. An example is shown in Figure~\ref{fig:skew_sln}.
	
	According to \cite[Eq.~(2.6)]{NT1998}, each column of height $i$
	contributes a factor $u+c$ to $P_i(u)$, where $c$ is the content
	of its bottom box. In $B_r$, the columns of height $i$ are precisely
	those with local column index $\nu_{i+1}^{(r)}<j\leq\nu_i^{(r)}$.
	Hence the block and full Drinfeld polynomials are
	\begin{equation}
		P_i^{(r)}(u)
		=\prod_{j=\nu_{i+1}^{(r)}+1}^{\nu_i^{(r)}}
		(u+s_r+j-i),
		\qquad
		P_i^\omega(u)=\prod_{r=1}^K P_i^{(r)}(u),
		\qquad i=1,\ldots,n-1.
		\label{eqn:drin_gen_snake/skew}
	\end{equation}
	Empty products equal $1$. The polynomials multiply because the
	blocks partition the columns without changing their heights or
	contents. The corresponding loop-weight monomial follows from
	\eqref{eqn:DrinPoly_snake}.
	
	\begin{rem}[Full columns and prime representations]
		\label{rem:full_columns}
		A full column of height $n$ has the unique semistandard filling
		$1,\ldots,n$ from top to bottom. In a skew diagram, a neighbouring
		column to the left ends at the same height or lower, whereas one
		to the right starts at the same height or higher.
		
		Consider the box with entry $a$ in the full column. Any box
		immediately to its left has at least $n-a$ boxes below it.
		Since entries increase strictly down columns and cannot exceed
		$n$, its entry is at most $a$. Similarly, a box immediately to
		the right has at least $a-1$ boxes above it, so its entry is at
		least $a$. The row inequalities across the full column are
		therefore already satisfied, and the tableaux on either side
		can be chosen independently. The tableau formula
		\cite[Theorem~3.4]{LuMukhin2021}, specialised to
		$Y(\mathfrak{gl}_n)$, therefore factorises the $q$-character into
		the contributions of the left part, the full column and the right
		part. Since skew modules are irreducible, injectivity of the
		$q$-character \cite[Lemma~2.8]{LuMukhin2021} gives the corresponding
		tensor-product decomposition. If the representations on both sides
		have dimension greater than one, the restriction to
		$Y(\mathfrak{sl}_n)$ is consequently not prime.
		
		At either outer edge, full columns contribute only one-dimensional
		determinant factors. In particular, full columns at the right edge
		can be removed while keeping the contents of all remaining boxes
		unchanged. As is also clear from \eqref{eqn:drin_gen_snake/skew},
		this leaves $P_1,\ldots,P_{n-1}$ unchanged. The omitted factors
		can be restored by multiplying the Lax operator by a scalar
		function $f(u)$. This changes the quantum determinant according to
		\begin{equation*}
			L(u)\longmapsto f(u)L(u),\qquad
			\detq L(u)\longmapsto
			\left(\prod_{j=0}^{n-1}f(u-j)\right)\detq L(u),
		\end{equation*}
		while leaving the $Y(\mathfrak{sl}_n)$-representation unchanged.
		Thus these factors remain part of the $\mathfrak{gl}_n$ data and
		are accounted for by the scalar normalisation of the Lax operator.
		
		These observations explain the restriction on full columns in
		the $\mathfrak{gl}_3$ diagram of Figure~\ref{fig:skew_sl3}:
		full columns separating two nontrivial parts exclude primality,
		and full columns at the
		right edge can be included through a scalar factor. The general
		construction below also applies to diagrams with full columns
		in the interior or at the right edge. $\odot$
	\end{rem}
	
	We next determine the lowest-weight eigenvalues. For each block,
	the product above telescopes to
	\begin{equation}
		\frac{P_i^{(r)}(u-1)}{P_i^{(r)}(u)}
		=\frac{u+s_r+\nu_{i+1}^{(r)}-i}
		{u+s_r+\nu_i^{(r)}-i}.
		\label{eqn:skew_block_Drinfeld_ratio}
	\end{equation}
	Writing $L_{i,i}(u)\xi=\ell_i(u)\xi$ on the lowest-weight
	vector $\xi$, \eqref{eqn:Shifted_L_ii_to_Drin_polys_direct} gives
	\begin{equation}
		\frac{\ell_{i+1}(u)}{\ell_i(u)}
		=\frac{P_i^\omega(u+i-1)}{P_i^\omega(u+i)}
		=\prod_{r=1}^K
		\frac{u+s_r+\nu_{i+1}^{(r)}}{u+s_r+\nu_i^{(r)}}.
		\label{eqn:skew_block_Lax_ratios}
	\end{equation}
	Thus each block contributes the linear factor
	$u+s_r+\nu_i^{(r)}$ to the $i$-th diagonal eigenvalue.
	These ratios determine the eigenvalues up to a common scalar
	factor. Set
	\begin{equation}
		d_i(u)\coloneqq\prod_{r=1}^K(u+s_r+\nu_i^{(r)}),
		\qquad
		g(u)\coloneqq\gcd(d_1(u),\ldots,d_n(u)),
		\label{eqn:skew_block_diagonal_products}
	\end{equation}
	where the GCD is monic. Choosing monic polynomial eigenvalues
	with GCD $1$, as in the $\mathfrak{gl}_3$ case, fixes this scalar
	normalisation and gives
	\begin{equation}
		L_{i,i}(u)\xi=\frac{d_i(u)}{g(u)}\,\xi,
		\qquad i=1,\ldots,n.
		\label{eqn:eigenval_gen_skew_lax_segment}
	\end{equation}
	Each $d_i$ has degree $K$, so these eigenvalues have degree
	$K-\deg g$. A common factor may occur even when all blocks are
	non-empty. Removing it leaves the Drinfeld polynomials unchanged.
	
	We now assign a linear oscillator Lax operator to each of these
	Young blocks. In the next subsection, Nazarov's fusion construction
	will fix their order of multiplication, and the tensor vacuum will
	generate the required skew representation. For the
	$21$-dimensional example below, we will also
	give the Gelfand--Tsetlin basis vectors explicitly as oscillator
	polynomials.

\subsection{Realisation through oscillators}

We construct the skew Lax operator by assigning a linear
oscillator Lax operator to each Young block of
Subsection~\ref{subsec:general_skew_Lax}. Nazarov's fusion
construction fixes the order of multiplication. In this order,
the tensor vacuum generates the desired irreducible module.

Using \eqref{eqn:lin_Lax} and a homomorphism from
$\mathfrak{gl}_n$ to the algebra $\mathcal A_n$ of
$n(n-1)/2$ oscillator pairs
$\{(\oa_{j,i},\oad_{i,j})\}_{1\leq i<j\leq n}$, with
\begin{equation}
	[\oa_{j,i},\oad_{k,l}]=\delta_{i,k}\delta_{j,l},
	\qquad [\oa_{j,i},\oa_{l,k}]=0,
	\qquad [\oad_{i,j},\oad_{k,l}]=0,
\end{equation}
we obtain the linear oscillator Lax operator \cite{FKT}
\begin{equation}
	\mathcal L_\lambda(u)=U^{-1}(u+D_\lambda)U,
	\label{eqn:lin_osc_Lax}
\end{equation}
where
\begin{equation}
	U=\begin{pmatrix}
		1 & -\oad_{1,2} & \cdots & -\oad_{1,n} \\
		0 & 1 & \ddots & \vdots \\
		\vdots & \ddots & \ddots & -\oad_{n-1,n} \\
		0 & \cdots & 0 & 1
	\end{pmatrix},
	\qquad
	D_\lambda=\begin{pmatrix}
		\lambda_1 & 0 & \cdots & 0\\[4pt]
		\widetilde\oa_{2,1} & \lambda_2-1 & \ddots & \vdots\\[4pt]
		\vdots & \ddots & \ddots & 0\\[4pt]
		\widetilde\oa_{n,1} & \cdots & \widetilde\oa_{n,n-1}
		& \lambda_n-(n-1)
	\end{pmatrix}
\end{equation}
and
\begin{equation}
	\widetilde\oa_{j,i}
	=-\oa_{j,i}+\sum_{k=j+1}^n\oad_{j,k}\oa_{k,i}.
\end{equation}
By \cite[Corollary~3.17(b)]{FKT}, the Fock vacuum generates
the finite-dimensional irreducible $\mathfrak{gl}_n$-module
$V_\lambda$ whenever
\begin{equation*}
	\lambda\in P^+
	=\{\lambda\in\C^n:
	\lambda_i-\lambda_{i+1}\in\Z_{\geq0},
	\ i=1,\ldots,n-1\}.
\end{equation*}
The common component $\lambda_n$ need not be an integer,
consistently with Subsection~\ref{subsec:eval_conventions}.

\medskip

\paragraph{\textit{Fusion of Young-diagram blocks.}} \leavevmode\par
For the block $B_r$ with shape $\nu^{(r)}$ and content shift
$s_r=p_r-q_r$, take
\begin{equation}
	\Lambda^{(r)}\coloneqq\nu^{(r)}+p_r(1,\ldots,1).
	\label{eqn:skew_block_oscillator_weights}
\end{equation}
These weights are dominant integral. On the module generated by
its vacuum, $\mathcal L_{\Lambda^{(r)}}(u-q_r)$ is the evaluation
Lax operator $L_{\nu^{(r)}}(u+s_r)$, by
\eqref{eqn:weight_shift_polynomial_Lax}. In particular, its diagonal
vacuum eigenvalues are
\begin{equation*}
	u-q_r+\Lambda_i^{(r)}=u+s_r+\nu_i^{(r)},
\end{equation*}
exactly the factors in \eqref{eqn:skew_block_diagonal_products}.
The common weight shift by $p_r$ is a convenient way to write this
evaluation module; it does not add boxes to the disjoint blocks
used in the fusion.

Nazarov's column tableau reads the columns from left to right and
each column from top to bottom. It therefore encounters our blocks
in the order $B_K,\ldots,B_1$. Using independent oscillators for
the different factors, we form
\begin{equation}
	\mathcal L(u)=
	\mathcal L_{\Lambda^{(K)}}^{[K]}(u-q_K)\cdots
	\mathcal L_{\Lambda^{(1)}}^{[1]}(u-q_1).
	\label{eqn:skew_ordered_oscillator_product}
\end{equation}
For example, $\lambda=(3,2)$ and $\mu=(1)$ give
\begin{equation*}
	(\nu^{(1)},p_1,q_1)=((2,1,0),1,0),
	\qquad
	(\nu^{(2)},p_2,q_2)=((1,0,0),0,1).
\end{equation*}
Thus the product for the $21$-dimensional example is
\begin{equation*}
	\mathcal L(u)=
	\mathcal L_{(1,0,0)}^{[2]}(u-1)
	\mathcal L_{(3,2,1)}^{[1]}(u).
\end{equation*}

\begin{thm}[The Oscillator-Snake Lax]
	\label{thm:Osc_Lax}
	Let $\omega=\lambda/\mu$ be a skew diagram with at most $n$ boxes
	in each column, and let $\mathcal L(u)$ be the ordered oscillator
	product \eqref{eqn:skew_ordered_oscillator_product}. The tensor
	vacuum
	\begin{equation*}
		\ket{\widetilde0}=\ket0_K\otimes\cdots\otimes\ket0_1
	\end{equation*}
	generates the finite-dimensional irreducible skew representation
	corresponding to $\omega$. Denote this module by $S$. Its oscillator Lax
	operator with the lowest-weight eigenvalues
	\eqref{eqn:eigenval_gen_skew_lax_segment} is
	\begin{equation*}
		L(u)=\left.\frac{1}{g(u)}\mathcal L(u)\right|_S,
	\end{equation*}
	where $g$ is defined in \eqref{eqn:skew_block_diagonal_products}.
	$\odot$
\end{thm}

\begin{proof}[Proof of Theorem~\ref{thm:Osc_Lax}]
	Let $\mathcal F_r$ be the Fock space of the $r$-th factor and set
	\begin{equation*}
		W_r=U(\mathfrak{gl}_n)\ket0_r\subset\mathcal F_r.
	\end{equation*}
	By \cite[Corollary~3.17(b)]{FKT}, $W_r$ is the finite-dimensional
	irreducible $\mathfrak{gl}_n$-module of highest weight
	$\Lambda^{(r)}$. The linear oscillator Lax has the form
	\begin{equation*}
		\mathcal L_{\Lambda^{(r)}}(u-q_r)
		=(u-q_r)I+\sum_{a,b=1}^n e_{ab}E_{ba}^{(r)}.
	\end{equation*}
	Dividing by $u-q_r$, as in
	\eqref{eqn:evaluation_RTT_convention}, gives the Yangian
	evaluation action. Its coefficients generate the same algebra
	as the $E_{ab}^{(r)}$, so $W_r$ is also an irreducible Yangian
	module generated by $\ket0_r$.
	
	The choice \eqref{eqn:skew_block_oscillator_weights} identifies
	this module with the straight block $B_r$. At the polynomial
	level,
	\begin{equation*}
		\left.\mathcal L_{\Lambda^{(r)}}(u-q_r)\right|_{W_r}
		\cong L_{\nu^{(r)}}(u+s_r).
	\end{equation*}
	For the matrices $T(u)$ of
	\eqref{eqn:evaluation_RTT_convention}, the scalar factor is
	\begin{equation*}
		T_{\Lambda^{(r)},q_r}(u)
		=\frac{u+s_r}{u-q_r}\,
		T_{\nu^{(r)},-s_r}(u).
	\end{equation*}
	Fusion of the ordinary Young diagram gives this evaluation
	representation after the same scalar normalisation; see
	\cite[Sections~1.3--1.5]{Nazarov2002}. The sign of $s_r$ follows
	from the convention change in
	Subsection~\ref{subsec:eval_conventions}. Multiplying a Lax
	operator by a scalar function does not change its invariant
	subspaces, and the identification maps the two lowest-weight
	vectors to each other up to a non-zero scalar.
	
	We next embed the full skew module in the tensor product of these
	blocks. Let $\Omega$ be Nazarov's column tableau of $\omega$ and
	$E_\Omega$ its regularised fusion operator on the fundamental
	tensor factors in tableau order. Its image is an irreducible
	skew module, with the scalar normalisation described in
	\cite[Proposition~2 and Theorem~2]{Nazarov2002}.
	Each $B_r$ occupies a consecutive interval in $\Omega$.
	Applying the suffix and prefix statements of
	\cite[Propositions~2.6 and~2.7]{NTBlockFusion2002} shows that
	\begin{equation*}
		E_\Omega=\widehat E_{B_r}X_r
	\end{equation*}
	for some operator $X_r$, where $\widehat E_{B_r}$ acts as the
	straight-block fusion operator on the factors of $B_r$ and as
	the identity on all other factors. The cited propositions give
	left and right divisibility of the regularised fusion elements
	themselves, so this factorisation holds at the prescribed fusion
	parameters.
	
	Let $H_r$ be the fundamental tensor factors belonging to $B_r$
	and $U_r=\operatorname{im}E_{B_r}\subset H_r$. The factorisation
	above puts $\operatorname{im}E_\Omega$ in
	$\operatorname{im}\widehat E_{B_r}$ for every $r$.
	Since the blocks are disjoint, intersecting these subspaces gives
	\begin{equation*}
		\operatorname{im}E_\Omega
		\subset U_K\otimes\cdots\otimes U_1.
	\end{equation*}
	Thus the fundamental factors within each Young block can indeed
	be fused first. After the scalar normalisations above, the
	identifications $U_r\cong W_r$ transfer the full skew module to
	an irreducible submodule
	\begin{equation*}
		S\subset W_K\otimes\cdots\otimes W_1
	\end{equation*}
	in precisely the order used in
	\eqref{eqn:skew_ordered_oscillator_product}.
	
	It remains to show that $S$ contains the tensor vacuum.
	Put $\gamma=\sum_{r=1}^K\Lambda^{(r)}$. By
	\eqref{eqn:drin_gen_snake/skew},
	\begin{equation*}
		\deg P_i^\omega
		=\sum_{r=1}^K\bigl(\nu_i^{(r)}-\nu_{i+1}^{(r)}\bigr)
		=\gamma_i-\gamma_{i+1},
		\qquad i=1,\ldots,n-1.
	\end{equation*}
	Thus $S$ and the tensor vacuum have the same
	$\mathfrak{sl}_n$ highest weight. The $\mathfrak{gl}_n$ centre
	acts by a fixed scalar on the whole tensor product, so the
	corresponding highest vector of $S$ has weight $\gamma$.
	Since each factor has a one-dimensional highest weight space,
	\begin{equation*}
		(W_K\otimes\cdots\otimes W_1)_\gamma
		=\C\ket{\widetilde0}.
	\end{equation*}
	Hence $\ket{\widetilde0}\in S$, and irreducibility gives
	\begin{equation*}
		Y(\mathfrak{gl}_n)\ket{\widetilde0}=S.
	\end{equation*}
	
	Finally, the ordered product of Lax operators realises the
	Yangian coproduct. Restricting it to $S$ gives the required
	oscillator realisation, with vacuum eigenvalues $d_i(u)$.
	Multiplying by $1/g(u)$ gives the normalisation
	\eqref{eqn:eigenval_gen_skew_lax_segment} without changing the
	subspace generated by the vacuum.
\end{proof}

\begin{prop}[The order of two factors]
\label{rem:reversed_fusion_order}
Suppose that Theorem~\ref{thm:Osc_Lax} involves two evaluation
factors. Write $M=V\otimes W$ in the prescribed order and
$M^{\mathrm{rev}}=W\otimes V$, including the spectral parameters
in $V$ and $W$. Let $S\subset M$ be the irreducible skew module
generated by the tensor vacuum $v$, and set $Q=M/S$.
If $Q\neq0$, the sequence
\begin{equation*}
    0\longrightarrow S\longrightarrow M\longrightarrow Q
    \longrightarrow0
\end{equation*}
is non-split. The reversed tensor vacuum $v^{\mathrm{rev}}$
generates all of $M^{\mathrm{rev}}$. $\odot$
\end{prop}

\begin{proof}
Let $r:M^{\mathrm{rev}}\to M$ be the non-zero regularised
$R$-matrix, including the exchange of tensor factors. Here
regularisation means taking the first non-zero Laurent coefficient
at the prescribed parameters. For an evaluation factor, the
regularised self-intertwiner is a non-zero scalar multiple of the
identity \cite[Proposition~4.7]{NTBlockFusion2002}.
The argument of \cite[Lemma~3.10, Corollary~3.11 and
Theorem~3.12]{KKKO2015}
therefore applies to these Yangian modules: it uses rigidity,
naturality and the tensor-product identities of the $R$-matrices,
together with this self-intertwiner property. Lemma~3.10 and the
non-vanishing argument of Corollary~3.11(i) hold in any rigid
abelian tensor category, so the compositions of regularised
intertwiners used in that argument remain non-zero here.
Taking leading Laurent coefficients in the rational Yangian
tensor-product identities therefore gives the required identities
up to non-zero scalars. It follows that
$M$ has a unique irreducible submodule and $M^{\mathrm{rev}}$
has a unique irreducible quotient, both given by $\operatorname{im}r$.
Since $S\subset M$ is irreducible, $\operatorname{im}r=S$.

We also obtain uniqueness of $r$ up to scalar. The extremal ordinary
weight space $M_\gamma=\C v$ from the proof of the theorem shows
that $S$ occurs with multiplicity one in $M$. The same holds in
$M^{\mathrm{rev}}$. The image of any non-zero map
$M^{\mathrm{rev}}\to M$ has simple head and socle isomorphic to
$S$. If this image were not simple, it would contain $S$ at least
twice as a composition factor. Hence its image is $S$, and the
map factors through the unique simple quotient of
$M^{\mathrm{rev}}$. Schur's lemma gives
\begin{equation*}
    \operatorname{Hom}_{Y(\mathfrak{gl}_n)}(M^{\mathrm{rev}},M)
    =\C r.
\end{equation*}

We use the transpose duality $D$ induced by
$T_{ij}(u)\mapsto T_{ji}(u)$. It fixes irreducible
finite-dimensional modules up to isomorphism and reverses tensor
products. Thus $D(M)\cong M^{\mathrm{rev}}$, and dualising gives
\begin{equation*}
    0\longrightarrow D(Q)\longrightarrow M^{\mathrm{rev}}
    \overset{q}{\longrightarrow}D(S)\longrightarrow0.
\end{equation*}
Composing $q$ with $D(S)\cong S$ and $S\hookrightarrow M$
gives a non-zero multiple of $r$. In particular,
$\ker q=\ker r$ is the unique maximal proper submodule of
$M^{\mathrm{rev}}$.

Suppose that $M=S\oplus Q$ with $Q\neq0$. Choose an irreducible
submodule $T\subset Q$ and write $j:T\hookrightarrow M$ for its
inclusion. Then
\begin{equation*}
    M^{\mathrm{rev}}\cong D(M)
    \overset{D(j)}{\longrightarrow}D(T)
    \cong T\overset{j}{\longrightarrow}M
\end{equation*}
is another non-zero intertwiner. Its image is $T\subset Q$,
so it cannot be proportional to $r$, whose image is $S$.
This contradicts uniqueness and proves non-splitting.

Finally, the identification $M^{\mathrm{rev}}\cong D(M)$
preserves ordinary weights. Since $M_\gamma=\C v$, the reversed
vacuum corresponds to a functional which is non-zero on $v$.
The map $q$ restricts this functional to $S$, so
$q(v^{\mathrm{rev}})\neq0$. Thus $v^{\mathrm{rev}}$ lies outside
the unique maximal proper submodule and generates all of
$M^{\mathrm{rev}}$.
\end{proof}

\begin{cor}\label{cor:Osc_snake_Lax_sl3}
	For the $\mathfrak{gl}_3$ skew diagram in
	Figure~\ref{fig:skew_sl3}, use the notation of
	\eqref{eqn:gl_3_skew_diag}, including its endpoint conventions
	$\mu_0=\lambda_1$ and $\mu_{k-1}=\lambda_{k+1}$.
	The latter is an auxiliary endpoint here, rather than a part
	of the partition $\mu$. The oscillator factors have weights
	\begin{equation*}
		\Lambda^{(r)}=(\mu_{r-1},\lambda_{r+1},\mu_r),
		\qquad q_r=r-1,\qquad r=1,\ldots,k-1.
	\end{equation*}
	Thus the ordered product is
	\begin{equation*}
		\mathcal L(u)=\overleftarrow{\prod}_{r=1}^{k-1}
		\mathcal L^{[r]}_{(\mu_{r-1},\lambda_{r+1},\mu_r)}(u-(r-1)),
	\end{equation*}
	with larger $r$ placed to the left.
	Its restriction to the module generated by the tensor vacuum
	is the skew Lax operator. Its diagonal vacuum eigenvalues are
	those in \eqref{eqn:gl_3_lax_lowest_EV}; a common scalar factor
	can be removed as in \eqref{eqn:eigenval_gen_skew_lax_segment}.
	$\odot$
\end{cor}

\subsection{Examples for \texorpdfstring{$\mathfrak{gl}_3$}{gl₃}}
In this section, we present some key examples of finite-dimensional Lax matrices whose oscillator bases we worked out explicitly. Two of them are quadratic in the spectral parameter. The first quadratic solution arises from the inverse relation for the linear Lax matrix \eqref{eqn:lin_Lax} in the adjoint (eight-dimensional) representation. The second has dimension $21$ and contains the Lie-algebra representations $15$ and $\bar{6}$.
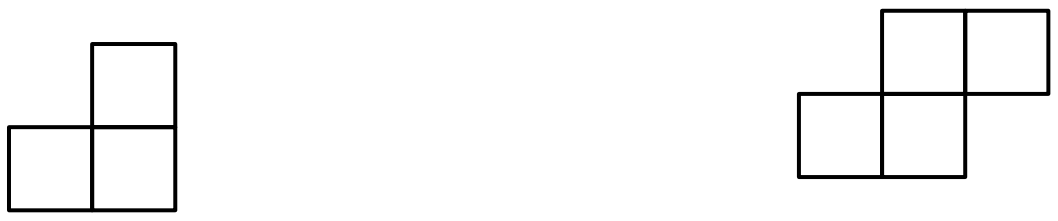
\begin{figure}[htbp]
	\centering
	
	\begin{tikzpicture}[
		x=0.55cm,
		y=0.55cm,
		line cap=round,
		line join=round
		]
		
		%
		%
		
		\begin{scope}[shift={(0,0)}]
			
			\draw[line width=0.75pt] (0,0) rectangle (1,1);
			\draw[line width=0.75pt] (1,0) rectangle (2,1);
			
			\draw[line width=0.75pt] (1,1) rectangle (2,2);
			
		\end{scope}

		%
		%
		
		\begin{scope}[shift={(9.5,0.4)}]
			
			\draw[line width=0.75pt] (0,0) rectangle (1,1);
			\draw[line width=0.75pt] (1,0) rectangle (2,1);
			
			\draw[line width=0.75pt] (1,1) rectangle (2,2);
			\draw[line width=0.75pt] (2,1) rectangle (3,2);
			
		\end{scope}
		
	\end{tikzpicture}
	
	\caption{
		Skew diagrams of the $8$-dimensional inverse adjoint representation (L)
		and the $21$-dimensional prime snake representation (R).
		Both correspond to quadratic Lax operators.
	}
	\label{fig:skew_diagrams_quadratic_lax}
	
\end{figure}
\subsubsection{Linear Lax}
We define
\begin{equation}
 U=\left(\begin{array}{ccc}
             1&-\oad_{1,2}&-\oad_{1,3}\\
             0&1&-\oad_{2,3}\\
             0&0&1
            \end{array}
\right)
\end{equation}
and consider the Lax matrix
\begin{equation}
 L_\lambda(u)=U^{-1}\left(\begin{array}{ccc}
             u+\lambda_1&0&0\\
             -\oa_{2,1}+\oad_{2,3}\oa_{3,1}&u+\lambda_2-1&0\\
             -\oa_{3,1}&-\oa_{3,2}&u+\lambda_3-2
            \end{array}
\right)U
\end{equation}
with
\begin{equation}
 (L_\lambda(u))_{aa}|0\rangle =(u+\lambda_a)|0\rangle
\end{equation}
The corresponding representation has highest weight $\lambda=(\lambda_1,\lambda_2,\lambda_3)$.

In particular, for the fundamental representation $\lambda=(1,0,0)$ we obtain
\begin{equation}
V=\operatorname{span}\{\oad_{1,3},\oad_{1,2},1\},
\end{equation}

For the antifundamental representation $\lambda=(1,1,0)$ we obtain
\begin{equation}
V=\operatorname{span}\{\oad_{1,2}\oad_{2,3}+\oad_{1,3},\oad_{2,3},1\},
\end{equation}
 
and for the adjoint representation $\lambda=(2,1,0)$ we obtain
\begin{equation}
V=\operatorname{span}\{\oad_{1,3}(\oad_{1,2}\oad_{2,3}+\oad_{1,3}),\oad_{1,2} (\oad_{1,2}\oad_{2,3}+\oad_{1,3}),\oad_{1,3}\,\oad_{2,3},\oad_{1,2}\,\oad_{2,3},\oad_{2,3},\oad_{1,3},\oad_{1,2},1\}
\label{eqn:adjoint_oscillator_basis}
\end{equation}

\subsubsection{Dual Lax}
Let $V^\vee$ denote the Hopf dual of a finite-dimensional Yangian
module $V$. Its action is defined by
\begin{equation*}
    \rho_{V^\vee}(a)=\rho_V(\mathsf S(a))^{t_V},
    \qquad
    T_{V^\vee}(u)=\bigl(T_V(u)^{-1}\bigr)^{t_V},
\end{equation*}
where $\mathsf S(T(u))=T(u)^{-1}$ is the antipode and $t_V$
denotes transposition in the quantum space. This is different
from the transpose duality $D$ in
Proposition~\ref{rem:reversed_fusion_order}.
The automorphism $T(u)\mapsto T(-u)^{-1}$
\cite[Section~1]{NT1998} also defines a Yangian action on $V$.
For the linear oscillator Lax, we obtain

\begin{equation}
\begin{split}
 {L}^{-1}_\lambda&(-u)=\rho(u) U^{-1}\\
 &\left(\begin{array}{ccc}
             (u-\lambda_2+1)(u-\lambda_3+2)&0&0\\
             (-\oa_{2,1}+\oad_{2,3}\oa_{3,1})(u-\lambda_3+2)&(u-\lambda_1)(u-\lambda_3+2)&0\\
             -\oa_{3,1}(u-\lambda_2+2)+(\oa_{2,1}-\oad_{2,3}\oa_{3,1})\oa_{3,2}&-\oa_{3,2}(u-\lambda_1)&(u-\lambda_1)(u-\lambda_2+1)
            \end{array}
\right)U\\
=&\rho(u)(u^2+u(e_{ab}J_{ba}-(\ell_1+\ell_2+\ell_3)\ID)+(e_{ab}J_{ba})^2-(\ell_1+\ell_2+\ell_3)e_{ab}J_{ba}+(\ell_1\ell_2+\ell_1 \ell_3+\ell_2\ell_3)\ID)\\
=&\rho(u)(u^2+u e_{ab}\bar J_{ba}+(e_{ab} \bar J_{ba})^2+(\ell_1+\ell_2+\ell_3)e_{ab}\bar J_{ba}+(\ell_1\ell_2+\ell_1 \ell_3+\ell_2\ell_3)\ID)
\end{split}
\end{equation}
where
\begin{equation}
 \rho(u)=(\lambda_1-u)^{-1}(\lambda_2-1-u)^{-1}(\lambda_3-2-u)^{-1}
\end{equation}
and
\begin{equation}
 \bar J_{ab}=J_{ab}-(\ell_1+\ell_2+\ell_3)\delta_{ab}
\end{equation}
with
\begin{align}
 (L^{-1}(-u))_{11}|0\rangle &=\rho(u)(u-\lambda_2+1)(u-\lambda_3+2)|0\rangle\\
 ( L^{-1}(-u))_{22}|0\rangle &=\rho(u)(u-\lambda_1+1)(u-\lambda_3+2)|0\rangle\\
  (L^{-1}(-u))_{33}|0\rangle &=\rho(u)(u-\lambda_1+1)(u-\lambda_2+2)|0\rangle
\end{align}
The inverse adjoint representation is realised on the same polynomial
space \eqref{eqn:adjoint_oscillator_basis} as the adjoint representation
$(2,1,0)$.
However, the adjoint and inverse adjoint are not isomorphic as Yangian representations. In both cases, the underlying $\mathfrak{sl}_3$-module is the same adjoint representation, but it admits two different non-factorisable irreducible Lax operators corresponding to distinct prime irreducible representations of the Yangian.

For $\lambda=(2,1,0)$, write $M=\sum_{a,b=1}^3e_{ab}J_{ba}$
and remove the scalar factor $\rho(u)$. This gives the polynomial
Lax operator
\begin{equation*}
    \widehat L(u)\coloneqq\rho(u)^{-1}L_{(2,1,0)}^{-1}(-u)
    =u^2\ID+uM+M^2-4\ID.
\end{equation*}
Its diagonal vacuum eigenvalues are monic and have GCD $1$, in
agreement with our convention.

To identify the Hopf dual, realise this adjoint representation
on the space of traceless $3\times3$ matrices, with
\begin{equation*}
    J_{ab}X=[e_{ab},X]+\delta_{ab}X,
    \qquad B(X,Y)=\operatorname{tr}(XY).
\end{equation*}
The term $\delta_{ab}X$ gives the $\mathfrak{gl}_3$ highest weight
$(2,1,0)$. The $\mathfrak{sl}_3$-invariant form $B$ identifies
$V^*$ with $V$. Denote its transpose by $\sharp$, so that
$B(AX,Y)=B(X,A^\sharp Y)$ for $A\in\operatorname{End}(V)$.
Then $J_{ab}^\sharp=-J_{ab}+2\delta_{ab}\ID$.
On Lax matrices, $\sharp$ acts only on the quantum-space entries;
in particular, $(M^2)^\sharp$ need not equal $(M^\sharp)^2$.
Using the $\mathfrak{gl}_3$ commutation relations, we obtain
\begin{equation*}
    M^\sharp=2\ID-M,
    \qquad
    (M^2)^\sharp=M^2-M+\ID.
\end{equation*}
With $T_V(u)=u^{-1}L_{(2,1,0)}(u)$, this gives
\begin{equation*}
    T_{V^\vee}(u)
    =\frac{\bigl(u^2\ID-uM+M^2-4\ID\bigr)^\sharp}{u^2-4}
    =\frac{\widehat L(u-1)}{u^2-4}.
\end{equation*}
Thus the inverse adjoint is related to the Hopf dual by the
displayed spectral translation and scalar normalisation.
The following subsection gives a fusion realisation of
$\widehat L(u)$.






\subsubsection{Product of two linear Lax matrices}
We fix $k=3$ in Corollary~\ref{cor:Osc_snake_Lax_sl3} so that the degree in the spectral parameter is $2$. Then we have

\begin{equation}
 \mathcal{L}(u)=
    L^{[2]}_{(\mu_{1},\lambda_3,\lambda_4)}(u-1) L^{[1]}_{(\lambda_1,\lambda_{2},\mu_{1})}(u)
\end{equation}
and it follows that
\begin{align}
		(\mathcal{L}(u))_{11}|0\rangle =& (u+\mu_{1}-1)(u+\lambda_1)|0\rangle\notag\\
		(\mathcal{L}(u))_{22}|0\rangle =& (u+\lambda_{3}-1)(u+\lambda_{2})|0\rangle\notag\\
		(\mathcal{L}(u))_{33}|0\rangle =& (u+\lambda_4-1)(u+\mu_{1})|0\rangle.\label{eqn:gl_3_lax_lowest_EV_deg2}
\end{align}
Then, for the inverse adjoint representation
\begin{equation}
 \lambda=(2,2,0,0)\,,\qquad \mu=(1),
\end{equation}
\begin{center}
	\begin{tikzpicture}[
		x=0.55cm,
		y=0.55cm,
		line cap=round,
		line join=round
		]
		
		\draw[line width=0.75pt] (0,0) rectangle (1,1);
		\draw[line width=0.75pt] (1,0) rectangle (2,1);
		
		\draw[line width=0.75pt] (1,1) rectangle (2,2);
		
	\end{tikzpicture}
\end{center}
we obtain the following basis, with the superscripts referring to the same
oscillator copies as in the product above:

\begin{equation}
\begin{pmatrix}
\oad_{1,2}^{[2]} (\oad_{1,2}^{[1]} \oad_{2,3}^{[1]} + \oad_{1,3}^{[1]}) \\
\oad_{1,2}^{[2]} \oad_{2,3}^{[1]} + \oad_{1,2}^{[1]} \oad_{2,3}^{[1]} + \oad_{1,3}^{[1]} \\
\oad_{1,2}^{[2]} \\
\oad_{1,3}^{[2]} (\oad_{1,2}^{[1]} \oad_{2,3}^{[1]} + \oad_{1,3}^{[1]}) \\
\oad_{2,3}^{[1]} \oad_{1,3}^{[2]} \\
\oad_{1,2}^{[1]} \oad_{2,3}^{[1]} + \oad_{1,3}^{[2]} + \oad_{1,3}^{[1]} \\
\oad_{2,3}^{[1]} \\
1
\end{pmatrix}
\end{equation}
generated by the vacuum $\ket{0}$.
We note that if we take the other order as
\begin{equation}
 \mathcal{L}(u)=L^{[1]}_{(2,2,1)}(u)
    L^{[2]}_{(1,0,0)}(u-1), 
\end{equation}
the vacuum generates the full nine-dimensional tensor product,
which contains a one-dimensional submodule.

\subsubsection{The 21-dimensional prime snake module}
\label{subsubsec:21dim_GT}
We now turn to the first quadratic example in our discussion which is not
obtained from the inversion relation. It is the $21$-dimensional prime snake
module associated with the skew diagram
\begin{equation}
 \lambda=(3,2,0,0)\,,\qquad \mu=(1).
\end{equation}
\begin{center}
	\begin{tikzpicture}[
		x=0.55cm,
		y=0.55cm,
		line cap=round,
		line join=round
		]
		
		\draw[line width=0.75pt] (0,0) rectangle (1,1);
		\draw[line width=0.75pt] (1,0) rectangle (2,1);
		
		\draw[line width=0.75pt] (1,1) rectangle (2,2);
		\draw[line width=0.75pt] (2,1) rectangle (3,2);
		
	\end{tikzpicture}
\end{center}
As a $\mathfrak{gl}_3$-module it decomposes as
$\mathbf{15}\oplus\overline{\mathbf{6}}$, whereas it is irreducible as a
Yangian module.

We obtain the product

\begin{equation}
 \mathcal{L}(u)=L^{[2]}_{(1,0,0)}(u-1) L^{[1]}_{(3,2,1)}(u)
 \label{eqn:21_GT_product}
\end{equation}
and it follows that
\begin{align}
 ( \mathcal{L}(u))_{11}|0\rangle &=u(u+3)|0\rangle\\
 (  \mathcal{L}(u))_{22}|0\rangle &=(u-1)(u+2)|0\rangle\\
  ( \mathcal{L}(u))_{33}|0\rangle &=(u-1)(u+1)|0\rangle.
\end{align}
In the convention of \eqref{eqn:Shifted_L_ii_to_Drin_polys_direct},
these eigenvalues give
\begin{equation}
    P_1(u)=(u-1)(u+2),\qquad P_2(u)=u,
    \qquad m_{21}=Y_{1,-2}Y_{2,1}Y_{1,4}.
    \label{eqn:21_Drinfeld_loop_origin}
\end{equation}
This is the common loop translation by $-2$ of the earlier example
$Y_{1,0}Y_{2,3}Y_{1,6}$, or equivalently a spectral translation by $1$.
Indeed, \eqref{eqn:snake_skew} places the earlier diagram in rows $0,1$;
translating it to rows $1,2$ gives the partition coordinates
$\lambda=(3,2,0,0)$, $\mu=(1)$ used here. Thus the relative snake
parameters are unchanged. The shift $\mathcal L(u+1)$ gives the parameters
used in the earlier $q$-character.

To make the oscillator realisation completely explicit, it is convenient to pass from the abstract oscillators to differential operators. For $n=3$ we set
\begin{equation}
	x_1=\oad_{1,2},\qquad x_2=\oad_{1,3},\qquad x_3=\oad_{2,3},
	\qquad
	\partial_1=\oa_{2,1},\qquad \partial_2=\oa_{3,1},\qquad \partial_3=\oa_{3,2}.
	\label{eqn:osc_diff_realisation}
\end{equation}
Then $[\partial_i,x_j]=\delta_{ij}$ and the oscillator algebra acts on $\C[x_1,x_2,x_3]$ by multiplication and differentiation. In these coordinates one has
\begin{equation}
	\widetilde\oa_{2,1}=-\partial_1+x_3\partial_2,
	\qquad
	\widetilde\oa_{3,1}=-\partial_2,
	\qquad
	\widetilde\oa_{3,2}=-\partial_3.
	\label{eqn:osc_diff_tilde}
\end{equation}
Hence the linear oscillator Lax matrix \eqref{eqn:lin_osc_Lax} for $\mathfrak{gl}_3$ can be written as
\begin{equation}
    \mathcal{L}_{\lambda}(u)=\bigl(L_{ab}(u)\bigr)_{a,b=1}^{3},
    \label{eqn:gl3_lin_osc_diff}
\end{equation}
where
\begin{align*}
    L_{11}(u)&=u+\lambda_1-x_1\partial_1-x_2\partial_2,\\
    L_{12}(u)&=(\lambda_2-\lambda_1)x_1
    +x_1^2\partial_1+x_1x_2\partial_2-(x_2+x_1x_3)\partial_3,\\
    L_{13}(u)&=-\lambda_1x_2-\lambda_2x_1x_3+\lambda_3(x_2+x_1x_3) +x_1x_2\partial_1+x_2^2\partial_2
    +x_3(x_2+x_1x_3)\partial_3,\\
    L_{21}(u)&=-\partial_1,\\
    L_{22}(u)&=u+\lambda_2+x_1\partial_1-x_3\partial_3,\\
    L_{23}(u)&=(\lambda_3-\lambda_2)x_3+x_2\partial_1+x_3^2\partial_3,\\
    L_{31}(u)&=-\partial_2,\\
    L_{32}(u)&=-\partial_3+x_1\partial_2,\\
    L_{33}(u)&=u+\lambda_3+x_2\partial_2+x_3\partial_3.
\end{align*}
For the product \eqref{eqn:21_GT_product} we therefore work on the polynomial ring
\begin{equation}
	\C[x_1^{[1]},x_2^{[1]},x_3^{[1]},x_1^{[2]},x_2^{[2]},x_3^{[2]}],
	\label{eqn:21_GT_poly_ring}
\end{equation}
where the superscripts $[1]$ and $[2]$ refer to the two tensor factors. The vacuum vector is the constant polynomial $\ket{0}=1$.
\medskip
\paragraph{\textit{An oscillator basis.}}
\leavevmode\par
To obtain an explicit basis, we start from the vacuum $\ket{0}=1$ and act with the strictly upper-triangular entries of the Lax matrix, i.e. with $\mathcal{L}_{ab}(u)$ for $a<b$. Since these entries are polynomials in the spectral parameter, we can equivalently act with their operator coefficients. We repeat this procedure on the newly generated states and discard linear dependencies. In this way the construction closes after $21$ linearly independent states. This gives a basis directly from the oscillator Lax matrix, but the basis is not yet adapted to the Gelfand--Tsetlin algebra.

The resulting oscillator basis is collected in Appendix~\ref{app:21_oscillator_basis}.

\medskip
\paragraph{\textit{The Gelfand--Tsetlin basis.}}
\leavevmode\par
To obtain a basis of common eigenvectors of the principal quantum
minors, we apply the ladder construction of
Subsection~\ref{subsec:GT_algebra}. On the vacuum, the eigenvalues are
\begin{align}
	a_1^{(0)}(u)&=u(u+3),\label{eqn:21_GT_vac_A1}\\
	a_2^{(0)}(u)&=(u-2)u(u+1)(u+3),\label{eqn:21_GT_vac_A2}\\
	a_3(u)&=(u-3)(u-2)(u-1)u(u+1)(u+3).\label{eqn:21_GT_vac_A3}
\end{align}
The last polynomial is the eigenvalue of the quantum determinant. Since the quantum determinant is central, $a_3(u)$ is the same for every state in the representation.

The zeros of $a_1^{(0)}(u)$ are $0$ and $-3$, while the zeros of $a_2^{(0)}(u)$ are $2,0,-1$ and $-3$. We can therefore evaluate the corresponding $B_m$ operators at these points. The non-zero actions on the vacuum are
\begin{align}
	B_1(0)\ket{0}
	&=-2\oad_{1,2}^{[2]},\label{eqn:21_GT_first_1}\\
	B_1(-3)\ket{0}
	&=3\oad_{1,2}^{[1]}+\oad_{1,2}^{[2]},\label{eqn:21_GT_first_2}\\
	B_2(-1)\ket{0}
	&=-6\oad_{2,3}^{[1]}.\label{eqn:21_GT_first_3}
\end{align}
Thus, the first level contains three states. In each case the corresponding zero is shifted by one according to \eqref{eqn:GT_root_shift}. We can now repeat the same procedure for the new states.

\medskip

\paragraph{\textit{Normalisation of the ladder operators.}}
 \leavevmode\par
With the polynomial normalisation used here, the ordinary prescription
$B_m(v)\ket{\Lambda}$ reaches only $16$ different Gelfand--Tsetlin
eigenlines. To obtain the remaining states, we remove a common scalar
factor from the ladder operators.

Let $V_{21}$ denote the $21$-dimensional module generated by the
vacuum. Direct calculation on its polynomial basis gives
\begin{equation*}
    A_2(u)|_{V_{21}}=u(u-2)\widehat A_2(u),
    \qquad
    B_2(u)|_{V_{21}}=u(u-2)\widehat B_2(u),
\end{equation*}
where $\widehat A_2(u)$ and $\widehat B_2(u)$ are polynomial
operators on $V_{21}$. They satisfy the same relation
\eqref{eqn:GT_AB_relation}, as explained in
Subsection~\ref{subsec:GT_algebra}. Evaluating and differentiating
the second identity at $u=0$ gives
\begin{equation*}
    B_2(0)|_{V_{21}}=0,
    \qquad
    B_2'(0)|_{V_{21}}=-2\widehat B_2(0).
\end{equation*}
If $a_2^\Lambda(u)$ has a double zero at $u=0$, then
$a_2^\Lambda(u)/[u(u-2)]$ has a simple zero there.
Thus, whenever $B_2'(0)\ket{\Lambda}\neq0$, it gives the ordinary
ladder step of the normalised operator, multiplied by $-2$.
We use the derivative form below to keep the original polynomial
normalisation.

A simple example already appears for
\begin{equation}
	\ket{\Lambda}
	=
	B_2(-1)B_1(0)\ket{0}
	=
	12\left(
	\oad_{2,3}^{[1]}\oad_{1,2}^{[2]}
	-\oad_{1,3}^{[2]}
	\right).
	\label{eqn:21_GT_confluent_parent}
\end{equation}
For this state, $a_2^\Lambda(u)$ has a double zero at $u=0$. Directly evaluating $B_2$ at this zero gives
\begin{equation}
	B_2(0)\ket{\Lambda}=0,
	\label{eqn:21_GT_confluent_zero}
\end{equation}
but its first derivative does not vanish,
\begin{equation}
	\ket{\Lambda'}\coloneqq
	\left.\frac{\mathrm d}{\mathrm dz}B_2(z)\ket{\Lambda}\right|_{z=0}
	=-144\,\oad_{2,3}^{[1]}\oad_{1,3}^{[2]}\neq0.
	\label{eqn:21_GT_confluent_derivative}
\end{equation}
The vector $\ket{\Lambda'}$ is again a Gelfand--Tsetlin eigenvector,
as explained in Subsection~\ref{subsec:GT_algebra}. Its eigenvalues are
\begin{equation}
	a_1^{\Lambda'}(u)=(u-1)(u+3),
	\qquad
	a_2^{\Lambda'}(u)=(u-2)(u-1)u(u+3),
\end{equation}
together with the common $a_3(u)$ in \eqref{eqn:21_GT_vac_A3}.

Repeating these steps and retaining the non-zero vectors gives the complete ladder. In this example there are five derivative steps, and all of them are of the form $B_2'(0)$. The numbers of states in the individual levels are
\begin{equation}
	1,\ 3,\ 4,\ 5,\ 4,\ 3,\ 1.
	\label{eqn:21_GT_level_dimensions}
\end{equation}
Their sum is $21$, as required. We also checked directly in the oscillator polynomial realisation that the $21$ states have rank $21$. Thus, they form a basis of the full module. Since snake modules are thin, every loop-weight space is one-dimensional, and by \eqref{eqn:GT_to_loop_weight} the Gelfand--Tsetlin eigenvalues $a_m^\Lambda(u)$ determine the corresponding loop weight. The full polynomial Gelfand--Tsetlin basis is collected in Table~\ref{tab:21_GT_basis}; we list the basis polynomials together with the eigenvalues of $A_1(u)$ and $A_2(u)$.

\begin{figure}[p]
	\centering
	\includegraphics[width=\linewidth]{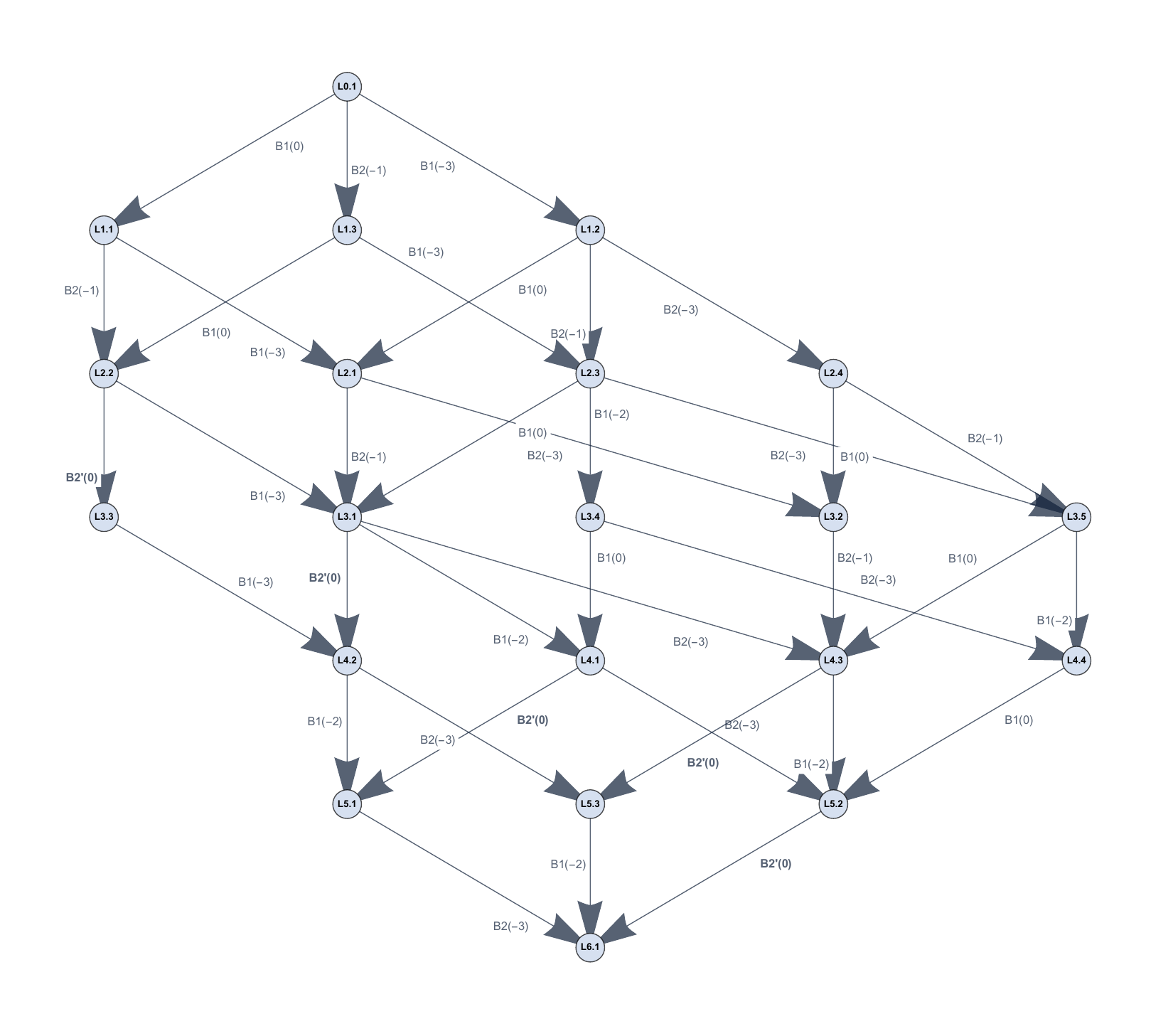}
	\caption{Gelfand--Tsetlin ladder of the $21$-dimensional prime snake module. The nodes are labelled by their level and their position inside the level. Ordinary edges correspond to the action of $B_m(v)$. The five bold edges $B_2'(0)$ are the additional first-derivative steps at double zeros in the polynomial normalisation used here. All arrows are understood up to a non-zero scalar. The level dimensions are $1,3,4,5,4,3,1$.}
	\label{fig:21_GT_ladder}
\end{figure}

Both basis constructions were implemented symbolically in \textit{Mathematica}. For the first basis, we represented the oscillators by multiplication and differential operators, applied the coefficients of the strictly upper-triangular entries of $\mathcal{L}(u)$ recursively to the vacuum, and kept a maximal linearly independent set of the resulting polynomials. For the Gelfand--Tsetlin basis, we constructed the quantum minors $A_m(u)$ and $B_m(u)$ from the same Lax matrix, determined the roots of the $A_m(u)$ eigenvalues, and applied the ladder operators and their derivatives recursively. Linear independence was checked by expanding the polynomial states in the monomial basis and computing the rank of the corresponding coefficient matrix. In this way, the explicit basis in Table~\ref{tab:21_GT_basis} and the ladder in Figure~\ref{fig:21_GT_ladder} were obtained directly from the same oscillator Lax matrix.

Reversing the order of the two factors gives
\begin{equation}
 \mathcal{L}(u)= L^{[1]}_{(3,2,1)}(u)L^{[2]}_{(1,0,0)}(u-1).
 \label{eqn:21_GT_reversed_product}
\end{equation}
In this example, a direct calculation of the intertwining equations
shows that the intertwiners between the two factor orders are unique
up to scalar in each direction and have ranks $21$ and $3$.
As explained in Proposition~\ref{rem:reversed_fusion_order}, the desired
$21$-dimensional skew module therefore appears as the quotient
in the reversed order, with an irreducible $3$-dimensional
submodule as kernel. The vacuum generates the entire
$24$-dimensional module in \eqref{eqn:21_GT_reversed_product}.

Figure~\ref{fig:21_GT_ladder} shows the ladder. Each node represents one Gelfand--Tsetlin basis vector. An ordinary edge labelled $B_m(v)$ means that applying $B_m(v)$ to the state at its beginning gives a non-zero scalar multiple of the state at its end. The five bold edges $B_2'(0)$ are the additional derivative steps described above.

\clearpage

\subsubsection{An alternating family}
\label{subsubsec:alternating_snakes}
We specialise the construction to the odd-length alternating family
in \cite[Eq.~(4.13)]{JuergensBoos2025}, with $n=2$ in the notation
of that paper and with a fixed common spectral origin. In our
loop-label convention, set
\begin{equation}
    m_k=\left(\prod_{r=0}^{k}Y_{2,6r-3}\right)
         \left(\prod_{r=0}^{k-1}Y_{1,6r}\right),
    \qquad S_k=V(m_k),\qquad k\geq0.
    \label{eqn:alternating_snake_monomial}
\end{equation}
The snake has $2k+1$ points, starting and ending at node $2$.
Consecutive labels differ by $3$, so it is minimal and prime.

\begin{cor}
\label{cor:alternating_oscillator_product}
Using independent oscillator copies, form
\begin{equation}
    \mathcal L_k(u)=
    \left(\overrightarrow{\prod}_{r=0}^{k-1}
      \mathcal L^{[r]}_{(2,1,0)}(u+3r-1)\right)
      \mathcal L^{[k]}_{(1,1,0)}(u+3k-1),
    \label{eqn:alternating_oscillator_product}
\end{equation}
where the factors in the ordered product are written with
increasing $r$ from left to right. The tensor vacuum generates an irreducible
$Y(\mathfrak{gl}_3)$-module whose $Y(\mathfrak{sl}_3)$-restriction
is $S_k$. Its Lax operator is the restriction of
$\mathcal L_k(u)$ to this submodule and has degree $k+1$.
Its Drinfeld polynomials are
\begin{equation}
    P_1^{(k)}(u)=\prod_{r=0}^{k-1}(u+3r),\qquad
    P_2^{(k)}(u)=\prod_{r=0}^{k}(u+3r-2).
    \label{eqn:alternating_Drinfeld}
\end{equation}
$\odot$
\end{cor}

\begin{proof}
Take the skew diagram with
\begin{equation*}
    \lambda_1=2k+1,\qquad
    \lambda_j=2k+5-2j\quad(2\leq j\leq k+2),\qquad
    \mu_j=2(k+1-j)\quad(1\leq j\leq k),
\end{equation*}
and extend both partitions by zeros. From left to right, its
Young blocks have shape $(2,1)$ and content shifts $3r-k$ for
$r=0,\ldots,k-1$, followed by a block of shape $(1,1)$ with
shift $2k$. Apply Theorem~\ref{thm:Osc_Lax} and then replace
$u$ by $u+k-1$. Absorbing the common weight shifts of the blocks
into their spectral parameters by \eqref{eqn:weight_shift_polynomial_Lax}
gives \eqref{eqn:alternating_oscillator_product}.
The block formula \eqref{eqn:drin_gen_snake/skew} gives
\eqref{eqn:alternating_Drinfeld}, and
\eqref{eqn:DrinPoly_snake} gives $m_k$. There are $k+1$ monic
linear factors, so the leading coefficient of the restricted
product is the identity and its degree is $k+1$.
\end{proof}

The diagonal vacuum eigenvalues are
\begin{align*}
    d_1^{(k)}(u)&=(u+3k)\prod_{r=0}^{k-1}(u+3r+1),\\
    d_2^{(k)}(u)&=\prod_{r=0}^{k}(u+3r),\qquad
    d_3^{(k)}(u)=\prod_{r=0}^{k}(u+3r-1).
\end{align*}
They have no common factor. The construction thus starts with
an antifundamental factor and adds $k$ adjoint factors, with all
relative shifts fixed by the snake. For $k=1$, the ordinary
$\mathfrak{sl}_3$ highest weight is $\omega_1+2\omega_2$;
the $21$-dimensional example above instead has highest weight
$2\omega_1+\omega_2$.

For an alternating snake of length $T$ with label spacing $3$,
the determinant in \eqref{eqn:general_snake_dimension} is
tridiagonal with diagonal entries $3$ and adjacent entries $1$.
Writing its dimension as $d_T$, expansion along the last row gives
\begin{equation}
    d_T=3d_{T-1}-d_{T-2},\qquad d_0=1,\quad d_1=3.
    \label{eqn:alternating_dimension_recurrence}
\end{equation}
In particular, $\dim S_k=d_{2k+1}$, with first values
$3,21,144$. This is a dimension recurrence; the oscillator
formula still requires restriction to the vacuum-generated
submodule. It does not establish the cancellation statement
of \cite[Conjecture~4.17]{JuergensBoos2025} for the snail operator.

\subsection{Applications to transfer matrices}
\label{subsec:transfer_applications}
\subsubsection{The 21-dimensional module as quantum space}
The resulting Lax matrix can be used to construct a homogeneous
periodic spin chain with local quantum space $V_{21}$ and fundamental
auxiliary space. For a chain of length $N$ without a boundary twist, define
\begin{equation}
    t_{\mathbf3\mid21}(u)
    =\operatorname{tr}_a\!\left(
       L_{aN}(u)\cdots L_{a1}(u)\right),
    \qquad a\cong\C^3,
    \label{eqn:21_quantum_transfer}
\end{equation}
where $L_{aj}(u)$ is the restriction of
\eqref{eqn:21_GT_product} to the $j$-th copy of $V_{21}$.
The $RLL$ relation implies that these transfer matrices commute.
In our spectral convention, the nested Bethe equations are
\cite[Section~2.6]{Arnaudon2005}
\begin{align}
    \left[\frac{u_k(u_k+3)}{(u_k-1)(u_k+2)}\right]^N
    &=\prod_{\substack{j=1\\j\neq k}}^{M_1}
      \frac{u_k-u_j+1}{u_k-u_j-1}
      \prod_{m=1}^{M_2}\frac{u_k-v_m-1}{u_k-v_m},
      \notag\\
    \left[\frac{v_m+2}{v_m+1}\right]^N
    &=\prod_{\substack{j=1\\j\neq m}}^{M_2}
      \frac{v_m-v_j+1}{v_m-v_j-1}
      \prod_{k=1}^{M_1}\frac{v_m-u_k}{v_m-u_k+1},
    \label{eqn:R21_BAE}
\end{align}
where $k=1,\ldots,M_1$ and $m=1,\ldots,M_2$ label the two sets
of Bethe roots. The driving terms are
$[P_1(u_k+1)/P_1(u_k)]^N$ and
$[P_2(v_m+2)/P_2(v_m+1)]^N$, with
$P_1(u)=(u-1)(u+2)$ and $P_2(u)=u$ as in
\eqref{eqn:21_Drinfeld_loop_origin}. These equations provide a
starting point for studying the spectrum of a chain whose local
space is irreducible under the Yangian but reducible under
$\mathfrak{gl}_3$.

\subsubsection{The 21-dimensional module as auxiliary space}
We can also use the module in the auxiliary space of a chain with
fundamental local spaces. Let $T_A(u)$ denote the auxiliary trace
of the monodromy with finite-dimensional auxiliary module $A$.
Its local $R$-operators are the Lax matrices with the roles of
the fundamental and representation spaces exchanged. Using one
tensor-product-compatible normalisation of the local $R$-operators,
we have
\begin{equation}
    T_{V\otimes W}(u)=T_V(u)T_W(u),\qquad
    T_M(u)=T_S(u)+T_Q(u)
    \quad\text{if }0\to S\to M\to Q\to0.
    \label{eqn:transfer_exact_sequence}
\end{equation}
The first identity follows from the coproduct. For the second,
choose a vector-space basis adapted to $S$. The monodromy is
block triangular, and its two diagonal blocks act on $S$ and $Q$.
Taking the auxiliary trace proves additivity, without requiring
the sequence to split.

In the example \eqref{eqn:21_GT_product}, the factors are the
evaluation modules with Lax matrices $L_{(1,0,0)}(u-1)$ and
$L_{(3,2,1)}(u)$. Their $q$-character product has dominant
monomials $m_{21}$ and $Y_{1,4}$. The latter belongs to the
three-dimensional quotient. Multiplicativity of the quantum
determinant gives \eqref{eqn:21_GT_vac_A3} on the whole tensor
product, hence also on the quotient. Its Lax operator is therefore
\begin{equation*}
    L_Q(u)=(u-1)L_{(3,2,2)}(u).
\end{equation*}
Indeed, this operator has Drinfeld polynomials $P_1(u)=u+2$,
$P_2(u)=1$ and the required quantum determinant; these data fix
the scalar normalisation as well as the irreducible module.
For a chain of length $N$, write $T_\lambda(u)$ for the transfer
matrix built from the linear evaluation Lax $L_\lambda(u)$.
Equation~\eqref{eqn:transfer_exact_sequence} gives the explicit identity
\begin{equation}
    T_{21}(u)=T_{(1,0,0)}(u-1)T_{(3,2,1)}(u)
       -(u-1)^N T_{(3,2,2)}(u).
    \label{eqn:21_transfer_fusion}
\end{equation}
Here $T_{21}$ uses \eqref{eqn:21_GT_product}, and $(u-1)^N$
comes from the $N$ local factors in $L_Q$. The identity also
holds with a common diagonal twist. For generic twist, each
evaluation transfer matrix on the right has the $3\times3$
$Q$-operator determinant of \cite[Eq.~(5.10)]{BFLMS}.

\section{Conclusion}
\label{sect:conclusion}

We constructed oscillator realisations of finite-dimensional skew
representations of $Y(\mathfrak{gl}_n)$, including prime snake
modules with non-factorisable Lax operators of higher degree.
The skew diagram determines the linear oscillator factors and
their spectral shifts through a decomposition into disjoint Young
blocks. With the order fixed by Nazarov's fusion, the tensor vacuum
generates precisely the irreducible skew module
(Theorem~\ref{thm:Osc_Lax}). Restricting the product to this
submodule gives its oscillator Lax operator.

The skew diagram gives the Drinfeld polynomials and diagonal
vacuum eigenvalues, while the path formula yields the dimension
determinant \eqref{eqn:general_snake_dimension} for arbitrary $n$.
The alternating family illustrates both the oscillator product
and the resulting dimension recurrence.

For the $21$-dimensional prime example, which is Yangian
irreducible but $\mathfrak{gl}_3$ reducible, we gave explicit
oscillator and Gelfand--Tsetlin bases. The principal quantum
minors determine the loop weights; removing scalar zeros explains
the derivative ladder steps. We also gave Bethe equations for
a chain with quantum space $V_{21}$ and the transfer-matrix
relation \eqref{eqn:21_transfer_fusion} for this auxiliary space.

The restriction to the vacuum-generated submodule remains the main
step to make more explicit. In the $21$-dimensional example,
the basis polynomials are independent of $x_3^{[2]}$:
the fundamental factor needs only two oscillator pairs, so five
pairs suffice for the product. For general skew modules, one can
ask how far this number can be reduced while keeping the entries
polynomial and of low degree in the oscillators; the realisations
of \cite{Frassek:2018try,Frassek:2020lky} provide candidates.
Relations on the skew submodule might also simplify the product
itself and lead to compact formulas or recursions under block fusion.

Explicit oscillator formulas for the fusion intertwiners could
help describe this restriction. Their images give the desired
submodules. For the non-split extensions of
Proposition~\ref{rem:reversed_fusion_order}, however, there is no
Yangian-equivariant projector onto the submodule, since such a
projector would split the sequence.

Oscillators may also be used in the auxiliary space.
Derkachov and Manashov constructed factorised $R$-operators
on pairs of oscillator representations for generic
$\mathfrak{sl}_n$ weights \cite{Derkachov:2008aq}.
Taking ordered products between two sets of Young blocks and
restricting to their vacuum modules could give oscillator formulas
for $R$-matrices between snake modules, for example on
$V_{21}\otimes V_{21}$. This requires a well-defined, non-zero
restriction; singular elementary factors at the fusion parameters
may need regularisation.

Beyond \eqref{eqn:21_transfer_fusion}, it remains to determine
which $Q$-operator expressions simplify for prime snake auxiliary
spaces. Evaluation determinants \cite[Eq.~(5.10)]{BFLMS} and the
known tableau, quantum Jacobi--Trudi and Wronskian-type formulas
\cite{Tsuboi1997,Tsuboi2010,LuMukhin2021} provide starting points.

The extended $T$-systems \cite{MY} and their quantum affine
cluster-algebra counterparts \cite{DuanLiLuo2019} offer a broader
framework for these relations. Likewise, the geometric descriptions
and shifted-Yangian constructions of \cite{Frassek:2018try,Frassek:2020lky}
may help identify oscillator coordinates in which the restriction
to the skew module becomes more explicit.

	\clearpage
	\appendix
	\section{The 21-dimensional module in polynomial form}
	\label{app:21dim_module}

	\subsection{An explicit oscillator basis}
	\label{app:21_oscillator_basis}
	We use the polynomial coordinates from \eqref{eqn:osc_diff_realisation}
	and the copy labels of \eqref{eqn:21_GT_product}: copy $[1]$ carries
	$V_{(3,2,1)}$ and copy $[2]$ carries $V_{(1,0,0)}$.
	In the realisation \eqref{eqn:21_GT_poly_ring}, the vacuum vector $1$
	generates a $21$-dimensional submodule. Successively applying the operator
	coefficients of the strictly upper-triangular entries $\mathcal{L}_{ij}(u)$,
	$i<j$, and retaining linearly independent states yields an oscillator basis.
	The following basis spans the same polynomial subspace as the
	Gelfand--Tsetlin basis in Table~\ref{tab:21_GT_basis}, but is not adapted
	to the commuting Gelfand--Tsetlin operators:
	\begingroup
	\setlength{\extrarowheight}{2pt}
	\begin{equation}
	\left(
	\begin{array}{c}
	x_2^{[1]} x_3^{[1]} x_2^{[2]} \\
	x_2^{[1]} x_2^{[2]} \left(x_1^{[1]} x_3^{[1]} + x_2^{[1]}\right) \\
	\frac{1}{3}\,\left(x_1^{[1]} x_2^{[1]} x_3^{[1]} - 3 x_1^{[1]} x_3^{[1]} x_2^{[2]} + \left(x_2^{[1]}\right)^{2}\right) \\
	- x_1^{[1]} x_2^{[2]} \left(x_1^{[1]} x_3^{[1]} + x_2^{[1]}\right) \\
	- x_1^{[1]} \left(x_1^{[1]} x_3^{[1]} + x_2^{[1]} + x_2^{[2]}\right) \\
	- x_3^{[1]} x_2^{[2]} \\
	-\frac{1}{3}\,\left(x_2^{[1]} + 3 x_2^{[2]}\right) \left(x_1^{[1]} x_3^{[1]} + x_2^{[1]}\right) \\
	-\frac{1}{2}\,\left(3 x_1^{[1]} x_3^{[1]} + 3 x_2^{[1]} + 2 x_2^{[2]}\right) \\
	- x_2^{[1]} \left(x_1^{[1]} x_3^{[1]} + x_2^{[1]} + x_3^{[1]} x_1^{[2]}\right) \\
	- x_2^{[1]} x_1^{[2]} \left(x_1^{[1]} x_3^{[1]} + x_2^{[1]}\right) \\
	\frac{1}{3}\,x_1^{[1]} \left(2 x_1^{[1]} x_3^{[1]} + 2 x_2^{[1]} + 3 x_3^{[1]} x_1^{[2]}\right) \\
	x_1^{[1]} x_1^{[2]} \left(x_1^{[1]} x_3^{[1]} + x_2^{[1]}\right) \\
	x_1^{[1]} x_1^{[2]} \\
	\frac{1}{2}\,\left(3 x_1^{[1]} x_3^{[1]} + 3 x_2^{[1]} + 2 x_3^{[1]} x_1^{[2]}\right) \\
	\frac{1}{3}\,\left(x_1^{[1]} + 3 x_1^{[2]}\right) \left(x_1^{[1]} x_3^{[1]} + x_2^{[1]}\right) \\
	x_1^{[2]} \\
	- x_2^{[1]} x_3^{[1]} \\
	\frac{1}{2}\,\left(x_1^{[1]} x_3^{[1]} - x_2^{[1]}\right) \\
	x_1^{[1]} \\
	x_3^{[1]} \\
	1 \\
	\end{array}
	\right)
	\label{eqn:21_oscillator_basis_appendix}
	\end{equation}
	\endgroup
	All entries are independent of $x_3^{[2]}$, consistently with the
	fundamental second factor, whose polynomial space is
	$\operatorname{span}\{1,x_1^{[2]},x_2^{[2]}\}$.

	\clearpage
	\subsection{Polynomial Gelfand--Tsetlin basis}
	\label{app:21_GT_basis}
	To keep the notation compact, we use the shorthand from \eqref{eqn:osc_diff_realisation}. Thus $x_1^{[s]},x_2^{[s]},x_3^{[s]}$ denote the multiplication operators in the $s$-th tensor factor. The table below lists the Gelfand--Tsetlin basis polynomials together with the eigenvalues of $A_1(u)$ and $A_2(u)$. The eigenvalue of $A_3(u)$ is the same for every state and equals
	\begin{equation}
		a_3(u)=(u-3)(u-2)(u-1)u(u+1)(u+3).
	\end{equation}

	{\footnotesize
	\renewcommand{\arraystretch}{1.02}
	\setlength{\tabcolsep}{3pt}
	\begin{longtable}{c c >{\raggedright\arraybackslash}p{8.2cm} >{\raggedright\arraybackslash}p{2.2cm} >{\raggedright\arraybackslash}p{3.0cm}}
	\caption{Polynomial Gelfand--Tsetlin basis of the $21$-dimensional module.}\label{tab:21_GT_basis}\\
	\hline
	Level & No. & $\ket{\Lambda}$ & $a_1^\Lambda(u)$ & $a_2^\Lambda(u)$\\
	\hline
	\endfirsthead
	\multicolumn{5}{c}{\tablename\ \thetable\ -- continued from previous page}\\
	\hline
	Level & No. & $\ket{\Lambda}$ & $a_1^\Lambda(u)$ & $a_2^\Lambda(u)$\\
	\hline
	\endhead
	\hline
	\multicolumn{5}{r}{Continued on next page}\\
	\endfoot
	\hline
	\endlastfoot
	0 & 1 & $1$ & $u(u+3)$ & $(u-2)u(u+1)(u+3)$\\
	1 & 1 & $-2x_1^{[2]}$ & $(u-1)(u+3)$ & $(u-2)u(u+1)(u+3)$\\
	1 & 2 & $3x_1^{[1]}+x_1^{[2]}$ & $u(u+2)$ & $(u-2)u(u+1)(u+3)$\\
	1 & 3 & $-6x_3^{[1]}$ & $u(u+3)$ & $(u-2)u^2(u+3)$\\
	2 & 1 & $-8x_1^{[1]}x_1^{[2]}$ & $(u-1)(u+2)$ & $(u-2)u(u+1)(u+3)$\\
	2 & 2 & $12(x_3^{[1]}x_1^{[2]}-x_2^{[2]})$ & $(u-1)(u+3)$ & $(u-2)u^2(u+3)$\\
	2 & 3 & $-3(3x_2^{[1]}+6x_1^{[1]}x_3^{[1]}+2x_3^{[1]}x_1^{[2]}+x_2^{[2]})$ & $u(u+2)$ & $(u-2)u^2(u+3)$\\
	2 & 4 & $-45(3x_2^{[1]}+x_2^{[2]})$ & $u(u+2)$ & $(u-2)u(u+1)(u+2)$\\
	3 & 1 & $24(x_2^{[1]}x_1^{[2]}+2x_1^{[1]}x_3^{[1]}x_1^{[2]}-x_1^{[1]}x_2^{[2]})$ & $(u-1)(u+2)$ & $(u-2)u^2(u+3)$\\
	3 & 2 & $120(3x_2^{[1]}x_1^{[2]}+x_1^{[1]}x_2^{[2]})$ & $(u-1)(u+2)$ & $(u-2)u(u+1)(u+2)$\\
	3 & 3 & $-144x_3^{[1]}x_2^{[2]}$ & $(u-1)(u+3)$ & $(u-2)(u-1)u(u+3)$\\
	3 & 4 & $-18(2x_1^{[1]}x_2^{[1]}+2(x_1^{[1]})^2x_3^{[1]}+x_2^{[1]}x_1^{[2]}+2x_1^{[1]}x_3^{[1]}x_1^{[2]}+x_1^{[1]}x_2^{[2]})$ & $u(u+1)$ & $(u-2)u^2(u+3)$\\
	3 & 5 & $135x_3^{[1]}(3x_2^{[1]}+x_2^{[2]})$ & $u(u+2)$ & $(u-2)u^2(u+2)$\\
	4 & 1 & $144x_1^{[1]}(x_2^{[1]}+x_1^{[1]}x_3^{[1]})x_1^{[2]}$ & $(u-1)(u+1)$ & $(u-2)u^2(u+3)$\\
	4 & 2 & $-192(x_2^{[1]}+2x_1^{[1]}x_3^{[1]})x_2^{[2]}$ & $(u-1)(u+2)$ & $(u-2)(u-1)u(u+3)$\\
	4 & 3 & $-360(3x_2^{[1]}x_3^{[1]}x_1^{[2]}-4x_2^{[1]}x_2^{[2]}+x_1^{[1]}x_3^{[1]}x_2^{[2]})$ & $(u-1)(u+2)$ & $(u-2)u^2(u+2)$\\
	4 & 4 & $810(2(x_2^{[1]})^2+2x_1^{[1]}x_2^{[1]}x_3^{[1]}+x_2^{[1]}x_3^{[1]}x_1^{[2]}+2x_2^{[1]}x_2^{[2]}+x_1^{[1]}x_3^{[1]}x_2^{[2]})$ & $u(u+1)$ & $(u-2)u^2(u+2)$\\
	5 & 1 & $-576x_1^{[1]}(x_2^{[1]}+x_1^{[1]}x_3^{[1]})x_2^{[2]}$ & $(u-1)(u+1)$ & $(u-2)(u-1)u(u+3)$\\
	5 & 2 & $-2160(x_2^{[1]}+x_1^{[1]}x_3^{[1]})(3x_2^{[1]}x_1^{[2]}+x_1^{[1]}x_2^{[2]})$ & $(u-1)(u+1)$ & $(u-2)u^2(u+2)$\\
	5 & 3 & $11520x_2^{[1]}x_3^{[1]}x_2^{[2]}$ & $(u-1)(u+2)$ & $(u-2)(u-1)u(u+2)$\\
	6 & 1 & $34560x_2^{[1]}(x_2^{[1]}+x_1^{[1]}x_3^{[1]})x_2^{[2]}$ & $(u-1)(u+1)$ & $(u-2)(u-1)u(u+2)$\\
	\end{longtable}
	}

	\clearpage
	\section{Determinant formula for skew representations}
	\label{app:skew_dimension_determinant}
	Figure~\ref{fig:dim_skew_sl3} displays the specialisation of
	\eqref{eqn:general_snake_dimension} to the $\mathfrak{gl}_3$ skew
	diagrams of \eqref{eqn:gl_3_skew_diag}.

	\begin{figure}[htbp]
		\centering
		
		\begin{tikzpicture}[
			x=0.43cm,
			y=0.43cm,
			line cap=round,
			line join=round
			]
			
			%
			%
			%
			
			\newcommand{\mentry}[3]{%
				\node at (#1,{-#2}) {$#3$};
			}

			
			\node[anchor=west] at (-2.95,2.35)
			{let};
			
			\node[anchor=west] at (-0.55,2.72)
			{$r_{2i-1}=\mu_{i-1}-\lambda_{i+1}$};
			
			\node[anchor=west] at (-0.55,1.88)
			{$r_{2i}=\lambda_{i+1}-\mu_i$};
			
			\node at (10.10,2.30)
			{$,$};
			
			\node[anchor=west] at (11.00,2.35)
			{then};
			
			\node[anchor=west] at (15.75,2.35)
			{$\dim(\lambda\backslash\mu)=$};

			
			\draw[line width=0.90pt]
			(-0.55,0.45)
			--
			(-0.55,-23.55);
			
			\draw[line width=0.90pt]
			(23.55,0.45)
			--
			(23.55,-23.55);

			%
			
			\node[font=\Large, anchor=east] at (-4.30,-9.65)
			{$=$};

			
			\mentry{0}{0}{3}
			\mentry{1}{0}{1}
			
			\mentry{0}{1}{3}
			\mentry{1}{1}{\cdot}
			\mentry{2}{1}{\cdot}
			
			\mentry{0}{2}{1}
			\mentry{1}{2}{\cdot}
			\mentry{2}{2}{\cdot}
			\mentry{3}{2}{\cdot}
			
			\mentry{1}{3}{\cdot}
			\mentry{2}{3}{\cdot}
			\mentry{3}{3}{\cdot}
			\mentry{4}{3}{\cdot}
			
			\mentry{2}{4}{\cdot}
			\mentry{3}{4}{\cdot}
			\mentry{4}{4}{\cdot}
			\mentry{5}{4}{1}
			
			\mentry{3}{5}{1}
			\mentry{4}{5}{3}
			\mentry{5}{5}{3}
			\mentry{6}{5}{1}

			
			\mentry{5}{6}{1}
			\mentry{6}{6}{3}
			\mentry{7}{6}{3}
			\mentry{8}{6}{1}
			
			\mentry{6}{7}{1}
			\mentry{7}{7}{\cdot}
			\mentry{8}{7}{\cdot}
			\mentry{9}{7}{\cdot}
			
			\mentry{7}{8}{\cdot}
			\mentry{8}{8}{\cdot}
			\mentry{9}{8}{\cdot}
			\mentry{10}{8}{\cdot}
			
			\mentry{8}{9}{\cdot}
			\mentry{9}{9}{\cdot}
			\mentry{10}{9}{\cdot}
			\mentry{11}{9}{1}
			
			\mentry{9}{10}{\cdot}
			\mentry{10}{10}{\cdot}
			\mentry{11}{10}{3}
			
			\mentry{10}{11}{1}
			\mentry{11}{11}{3}
			\mentry{12}{11}{1}
			
			\mentry{11}{12}{1}

			
			\mentry{13}{13}{\cdot}
			\mentry{14}{14}{\cdot}
			\mentry{15}{15}{\cdot}

			
			\mentry{18}{17}{1}
			
			\mentry{17}{18}{1}
			\mentry{18}{18}{3}
			\mentry{19}{18}{3}
			\mentry{20}{18}{1}
			
			\mentry{18}{19}{1}
			\mentry{19}{19}{\cdot}
			\mentry{20}{19}{\cdot}
			\mentry{21}{19}{\cdot}
			
			\mentry{19}{20}{\cdot}
			\mentry{20}{20}{\cdot}
			\mentry{21}{20}{\cdot}
			\mentry{22}{20}{\cdot}
			
			\mentry{20}{21}{\cdot}
			\mentry{21}{21}{\cdot}
			\mentry{22}{21}{\cdot}
			\mentry{23}{21}{1}
			
			\mentry{21}{22}{\cdot}
			\mentry{22}{22}{\cdot}
			\mentry{23}{22}{3}
			
			\mentry{22}{23}{1}
			\mentry{23}{23}{3}

			
			\draw[
			decorate,
			decoration={brace,amplitude=5pt,mirror},
			line width=0.65pt
			]
			(-1.55,0.35)
			--
			(-1.55,-5.35);
			
			\node[anchor=east]
			at (-1.95,-2.50)
			{$r_1$};

			%
			
			\draw[
			decorate,
			decoration={brace,amplitude=5pt,mirror},
			line width=0.65pt
			]
			(-1.55,-5.65)
			--
			(-1.55,-11.35);
			
			\node[anchor=east]
			at (-1.95,-8.50)
			{$r_2$};

			%
			
			\node at (-1.95,-13.20) {$\cdot$};
			\node at (-1.95,-14.00) {$\cdot$};
			\node at (-1.95,-14.80) {$\cdot$};

			%
			
			\draw[
			decorate,
			decoration={brace,amplitude=5pt,mirror},
			line width=0.65pt
			]
			(-1.55,-17.65)
			--
			(-1.55,-23.35);
			
			\node[anchor=east]
			at (-1.95,-20.50)
			{$r_{2k-2}$};

		\end{tikzpicture}
		
		\caption{
			Determinant formula for the dimension of the skew representation
			$\lambda\backslash\mu$. The definitions of $r_{2i-1}$ and
			$r_{2i}$ use $i=1,\ldots,k-1$; blocks of size zero are omitted.
		}
		\label{fig:dim_skew_sl3}
	\end{figure}

	\clearpage
	\bibliographystyle{amsplain-doi}
	\bibliography{references}

@article{Molev00,
  author        = {Molev, A.},
  title         = {{Irreducibility criterion for tensor products of {Yangian} evaluation modules}},
  journal       = {Duke Mathematical Journal},
  volume        = {112},
  number        = {2},
  pages         = {307--341},
  year          = {2002},
  doi           = {10.1215/S0012-9074-02-11224-1},
  eprint        = {math/0009183},
  archivePrefix = {arXiv}
}

@article{HL,
  author        = {Hernandez, D. and Leclerc, B.},
  title         = {{Cluster algebras and quantum affine algebras}},
  journal       = {Duke Math. J.},
  volume        = {154},
  number        = {2},
  pages         = {265--341},
  year          = {2010},
  doi           = {10.1215/00127094-2010-040},
  eprint        = {0903.1452},
  archivePrefix = {arXiv}
}

@article{FM,
  author  = {Frenkel, E. and Mukhin, E.},
  title   = {{Combinatorics of q-characters of finite-dimensional representations of quantum affine algebras}},
  journal = {Commun. Math. Phys.},
  volume  = {216},
  number  = {1},
  pages   = {23--57},
  year    = {2001},
  doi     = {10.1007/s002200000323}
}

@incollection{FR,
  author        = {Frenkel, E. and Reshetikhin, N.},
  title         = {{The {$q$}-characters of representations of quantum affine algebras and deformations of {$W$}-algebras}},
  booktitle     = {Recent Developments in Quantum Affine Algebras and Related Topics},
  series        = {Contemporary Mathematics},
  volume        = {248},
  pages         = {163--205},
  year          = {1999},
  publisher     = {American Mathematical Society},
  address       = {Providence, RI},
  doi           = {10.1090/conm/248/03823},
  eprint        = {math/9810055},
  archivePrefix = {arXiv}
}

@article{MY2,
  author  = {Mukhin, E. and Young, C. A. S.},
  title   = {{Path description of type B q-characters}},
  journal = {Advances in Mathematics},
  volume  = {231},
  number  = {2},
  pages   = {1119--1150},
  year    = {2012},
  doi     = {10.1016/j.aim.2012.06.012}
}

@article{MY,
  author        = {Mukhin, E. and Young, C. A. S.},
  title         = {{Extended T-systems}},
  journal       = {Sel. Math. New Ser.},
  volume        = {18},
  number        = {3},
  pages         = {591--631},
  year          = {2012},
  doi           = {10.1007/s00029-011-0083-x},
  eprint        = {1104.3094},
  archivePrefix = {arXiv}
}

@book{molev2007yangians,
  author    = {Molev, A.},
  title     = {{Yangians and Classical Lie Algebras}},
  series    = {Mathematical Surveys and Monographs},
  volume    = {143},
  year      = {2007},
  publisher = {American Mathematical Society},
  doi       = {10.1090/surv/143},
  isbn      = {9780821843741},
  url       = {https://doi.org/10.1090/surv/143}
}

@article{Drinfeld1988,
  author  = {Drinfeld, V. G.},
  title   = {{A new realization of Yangians and of quantum affine algebras}},
  journal = {Soviet Math. Dokl.},
  volume  = {36},
  number  = {2},
  pages   = {212--216},
  year    = {1988},
  url     = {https://www.mathnet.ru/eng/dan7988}
}

@article{N,
  author  = {Nakajima, H.},
  title   = {{t-analogs of q-characters of Kirillov-Reshetikhin modules of quantum affine algebras}},
  journal = {Represent. Theory},
  volume  = {7},
  pages   = {259--274},
  year    = {2003},
  doi     = {10.1090/S1088-4165-03-00164-X}
}

@article{GT,
  author  = {Gautam, S. and Toledano Laredo, V.},
  title   = {{Yangians, quantum loop algebras, and abelian difference equations}},
  journal = {J. Amer. Math. Soc.},
  volume  = {29},
  number  = {3},
  pages   = {775--824},
  year    = {2016},
  doi     = {10.1090/jams/851}
}

@article{C2002,
  author  = {Chari, V.},
  title   = {{Braid group actions and tensor products}},
  journal = {Int. Math. Res. Not.},
  number  = {7},
  pages   = {357--382},
  year    = {2002},
  doi     = {10.1155/S107379280210612X}
}

@book{CPBook,
  author    = {Chari, V. and Pressley, A.},
  title     = {{A guide to quantum groups}},
  year      = {1994},
  publisher = {Cambridge University Press},
  url       = {https://assets.cambridge.org/97805215/58846/frontmatter/9780521558846_frontmatter.pdf}
}

@article{NT1998,
  author  = {Nazarov, M. and Tarasov, V.},
  title   = {{Representations of Yangians with Gelfand-Zetlin Bases}},
  journal = {J. Reine Angew. Math.},
  volume  = {496},
  pages   = {181--212},
  year    = {1998},
  doi     = {10.1515/crll.1998.029}
}

@article{FKT,
  author  = {Frassek, R. and Karpov, I. and Tsymbaliuk, A.},
  title   = {{Transfer Matrices of Rational Spin Chains via Novel BGG-Type Resolutions}},
  journal = {Commun. Math. Phys.},
  volume  = {400},
  number  = {1},
  pages   = {1--82},
  year    = {2023},
  doi     = {10.1007/s00220-022-04620-6}
}

@misc{JuergensMaster,
  author        = {J{\"u}rgens, H.},
  title         = {{On the structure of tensor products of the finite-dimensional representations of quantum affine $\mathfrak{sl}_n$}},
  year          = {2025},
  doi           = {10.48550/arXiv.2501.14801},
  eprint        = {2501.14801},
  archivePrefix = {arXiv},
  url           = {https://arxiv.org/abs/2501.14801},
  note          = {Master's thesis; arXiv version}
}

@inproceedings{Nazarov2002,
  author        = {Nazarov, M.},
  title         = {{Representations of Yangians associated with skew Young diagrams}},
  booktitle     = {Proceedings of the International Congress of Mathematicians, Beijing 2002},
  volume        = {2},
  pages         = {643--654},
  year          = {2002},
  doi           = {10.48550/arXiv.math/0209129},
  eprint        = {math/0209129},
  archivePrefix = {arXiv},
  url           = {https://arxiv.org/abs/math/0209129},
  note          = {DOI refers to the arXiv version}
}

@article{BrundanKleshchev2005,
  author        = {Brundan, J. and Kleshchev, A.},
  title         = {{Parabolic presentations of the {Yangian} {$Y(\mathfrak{gl}_n)$}}},
  journal       = {Communications in Mathematical Physics},
  volume        = {254},
  number        = {1},
  pages         = {191--220},
  year          = {2005},
  doi           = {10.1007/s00220-004-1249-6},
  eprint        = {math/0407011},
  archivePrefix = {arXiv}
}

@article{NTBlockFusion2002,
  author        = {Nazarov, M. and Tarasov, V.},
  title         = {{On irreducibility of tensor products of {Yangian} modules associated with skew {Young} diagrams}},
  journal       = {Duke Mathematical Journal},
  volume        = {112},
  number        = {2},
  pages         = {343--378},
  year          = {2002},
  doi           = {10.1215/S0012-9074-02-11225-3},
  eprint        = {math/0012039},
  archivePrefix = {arXiv},
  url           = {https://arxiv.org/abs/math/0012039}
}

@article{LuMukhin2021,
  author        = {Lu, K. and Mukhin, E.},
  title         = {{{Jacobi--Trudi} Identity and {Drinfeld} Functor for Super {Yangian}}},
  journal       = {International Mathematics Research Notices},
  volume        = {2021},
  number        = {21},
  pages         = {16751--16810},
  year          = {2021},
  doi           = {10.1093/imrn/rnab023},
  eprint        = {2007.15573},
  archivePrefix = {arXiv}
}

@article{KKKO2015,
  author        = {Kang, S.-J. and Kashiwara, M. and Kim, M. and Oh, S.-J.},
  title         = {{Simplicity of heads and socles of tensor products}},
  journal       = {Compositio Mathematica},
  volume        = {151},
  number        = {2},
  pages         = {377--396},
  year          = {2015},
  doi           = {10.1112/S0010437X14007799},
  eprint        = {1404.4125},
  archivePrefix = {arXiv},
  primaryClass  = {math.RT}
}

@article{DuanLiLuo2019,
  author        = {Duan, B. and Li, J.-R. and Luo, Y.-F.},
  title         = {{Cluster algebras and snake modules}},
  journal       = {Journal of Algebra},
  volume        = {519},
  pages         = {325--377},
  year          = {2019},
  doi           = {10.1016/j.jalgebra.2018.10.027},
  eprint        = {1508.03467},
  archivePrefix = {arXiv}
}

@article{Frassek:2018try,
  author        = {Frassek, R. and Pestun, V.},
  title         = {{A Family of {$\mathrm{GL}_r$} Multiplicative Higgs Bundles on Rational Base}},
  journal       = {SIGMA},
  volume        = {15},
  pages         = {031},
  year          = {2019},
  doi           = {10.3842/SIGMA.2019.031},
  eprint        = {1808.00799},
  archivePrefix = {arXiv},
  url           = {https://sigma-journal.com/2019/031/}
}

@article{Frassek:2020lky,
  author        = {Frassek, R. and Pestun, V. and Tsymbaliuk, A.},
  title         = {{Lax matrices from antidominantly shifted {Yangians} and quantum affine algebras: {A}-type}},
  journal       = {Advances in Mathematics},
  volume        = {401},
  pages         = {108283},
  year          = {2022},
  doi           = {10.1016/j.aim.2022.108283},
  eprint        = {2001.04929},
  archivePrefix = {arXiv},
  url           = {https://arxiv.org/abs/2001.04929}
}

@article{Derkachov:2008aq,
  author        = {Derkachov, S. E. and Manashov, A. N.},
  title         = {{Factorization of {$R$}-matrix and {Baxter} {$Q$}-operators for generic {$\mathfrak{sl}(N)$} spin chains}},
  journal       = {Journal of Physics A: Mathematical and Theoretical},
  volume        = {42},
  number        = {7},
  pages         = {075204},
  year          = {2009},
  doi           = {10.1088/1751-8113/42/7/075204},
  eprint        = {0809.2050},
  archivePrefix = {arXiv}
}

@article{BFLMS,
  author        = {Bazhanov, V. V. and Frassek, R. and {\L}ukowski, T. and Meneghelli, C. and Staudacher, M.},
  title         = {{{Baxter} {$Q$}-operators and representations of {Yangians}}},
  journal       = {Nuclear Physics B},
  volume        = {850},
  number        = {1},
  pages         = {148--174},
  year          = {2011},
  doi           = {10.1016/j.nuclphysb.2011.04.006},
  eprint        = {1010.3699},
  archivePrefix = {arXiv},
  url           = {https://arxiv.org/abs/1010.3699}
}

@article{Green:1971rp,
  author  = {Green, H. S.},
  title   = {{Characteristic identities for generators of gl(n), o(n) and sp(n)}},
  journal = {J. Math. Phys.},
  volume  = {12},
  number  = {10},
  pages   = {2106--2113},
  year    = {1971},
  doi     = {10.1063/1.1665508}
}

@article{GesselViennot1985,
  author  = {Gessel, I. and Viennot, G.},
  title   = {{Binomial determinants, paths, and hook length formulae}},
  journal = {Advances in Mathematics},
  volume  = {58},
  number  = {3},
  pages   = {300--321},
  year    = {1985},
  doi     = {10.1016/0001-8708(85)90121-5}
}

@article{Tsuboi1997,
  author        = {Tsuboi, Z.},
  title         = {{Analytic {Bethe} ansatz and functional equations for {Lie} superalgebra {$\mathfrak{sl}(r+1|s+1)$}}},
  journal       = {Journal of Physics A: Mathematical and General},
  volume        = {30},
  number        = {22},
  pages         = {7975--7991},
  year          = {1997},
  doi           = {10.1088/0305-4470/30/22/031},
  eprint        = {0911.5386},
  archivePrefix = {arXiv}
}

@article{Tsuboi2010,
  author        = {Tsuboi, Z.},
  title         = {{Solutions of the {$T$}-system and {Baxter} equations for supersymmetric spin chains}},
  journal       = {Nuclear Physics B},
  volume        = {826},
  number        = {3},
  pages         = {399--455},
  year          = {2010},
  doi           = {10.1016/j.nuclphysb.2009.08.009},
  eprint        = {0906.2039},
  archivePrefix = {arXiv}
}

@article{JuergensBoos2025,
  author        = {J{\"u}rgens, H. and Boos, H.},
  title         = {{On the properties of the density matrix of the {$\mathfrak{sl}_{n+1}$}-invariant model}},
  journal       = {Journal of Statistical Mechanics: Theory and Experiment},
  number        = {3},
  pages         = {033104},
  year          = {2025},
  doi           = {10.1088/1742-5468/adbdcb},
  eprint        = {2308.15439},
  archivePrefix = {arXiv}
}

@article{KhoroshkinNazarov2012,
  author        = {Khoroshkin, S. and Nazarov, M.},
  title         = {{{Mickelsson} algebras and representations of {Yangians}}},
  journal       = {Transactions of the American Mathematical Society},
  volume        = {364},
  number        = {3},
  pages         = {1293--1367},
  year          = {2012},
  doi           = {10.1090/S0002-9947-2011-05367-5},
  eprint        = {0912.1101},
  archivePrefix = {arXiv},
  primaryClass  = {math.RT}
}

@article{KhoroshkinNazarovShapiro2014,
  author        = {Khoroshkin, S. and Nazarov, M. and Shapiro, A.},
  title         = {{Rational and polynomial representations of {Yangians}}},
  journal       = {Journal of Algebra},
  volume        = {418},
  pages         = {265--291},
  year          = {2014},
  doi           = {10.1016/j.jalgebra.2014.06.038},
  eprint        = {1408.4924},
  archivePrefix = {arXiv},
  primaryClass  = {math.RT}
}

@article{vanLeeuwen2006,
  author        = {van Leeuwen, M. A. A.},
  title         = {{{Schur} functions and alternating sums}},
  journal       = {The Electronic Journal of Combinatorics},
  volume        = {11},
  number        = {2},
  pages         = {A5},
  year          = {2006},
  doi           = {10.37236/1897},
  eprint        = {math/0602357},
  archivePrefix = {arXiv}
}

@article{Arnaudon2005,
  author        = {Arnaudon, D. and Cramp{\'e}, N. and Doikou, A. and Frappat, L. and Ragoucy, E.},
  title         = {{Analytical {Bethe} Ansatz for closed and open {$\mathfrak{gl}(n)$}-spin chains in any representation}},
  journal       = {Journal of Statistical Mechanics: Theory and Experiment},
  number        = {2},
  pages         = {P02007},
  year          = {2005},
  doi           = {10.1088/1742-5468/2005/02/P02007},
  eprint        = {math-ph/0411021},
  archivePrefix = {arXiv}
}

\end{document}